\documentclass[twocolumn,floatfix,superscriptaddress,aps,prl,noeprint]{revtex4-2}

\usepackage{dsfont} 

\usepackage{bm} 

\usepackage{bbold} 
\usepackage{times}
\usepackage{graphicx,braket}
\usepackage{hyperref}
\hypersetup{colorlinks=true,linkcolor=blue,citecolor=blue}
\usepackage{amsmath,amssymb,amsfonts}
\usepackage{wrapfig}

\usepackage{siunitx}

\usepackage{cleveref}
\crefname{equation}{Eqs.}{Eqs.}
\Crefname{equation}{Equation}{Equations}
\crefrangelabelformat{equation}{(#3#1#4--#5#2#6)}
\crefmultiformat{equation}{Eqs. (#2#1#3}{, #2#1#3)}{#2#1#3}{#2#1#3}
\Crefmultiformat{equation}{Equations (#2#1#3}{, #2#1#3)}{#2#1#3}{#2#1#3}

\usepackage{booktabs}
\usepackage{mathtools}

\newcommand{\bfop}[1]{\hat{\bf #1}}

\newcommand{\sz}{\hat \sigma_z}

\newcommand{\sx}{\hat \sigma_x}
\newcommand{\spol}{\hat\sigma_p}

\newcommand{\sop}{\hat\sigma}
\newcommand{\sdop}{\hat\sigma^\dagger}

\newcommand{\aop}{\hat a}
\newcommand{\adop}{\hat a ^\dagger}

\renewcommand{\eqref}[1]{\mbox{Eq.~(\ref{#1})}}

\newcommand{\dm}{\hat\rho}

\newcommand{\nr}{n_\mathrm{R}}
\newcommand{\nl}{n_\mathrm{L}}
\newcommand{\gr}{g_\mathrm{R}}
\newcommand{\gl}{g_\mathrm{L}}
\newcommand{\omr}{\omega_\mathrm{R}}
\newcommand{\oml}{\omega_\mathrm{L}}

\newcommand{\be}{\begin{equation}}
\newcommand{\ee}{\end{equation}}
\newcommand{\bea}{\begin{eqnarray}}
\newcommand{\eea}{\end{eqnarray}}

\newcommand{\ham}{\hat H}

\begin{document}

\flushbottom
\title{Spontaneous Scattering of Raman Photons from Cavity-QED Systems in the Ultrastrong Coupling Regime}

\author{Vincenzo Macrì}
\affiliation{Theoretical Quantum Physics Laboratory, RIKEN Cluster for Pioneering Research, Wakoshi, Saitama 351-0198, Japan}

\author{Alberto Mercurio}
\affiliation{Dipartimento di Scienze Matematiche e Informatiche,
Scienze Fisiche e Scienze della Terra, Universit{à} di Messina, I-98166 Messina, Italy}
\author{Franco Nori}
\affiliation{Theoretical Quantum Physics Laboratory, RIKEN Cluster for Pioneering Research, Wakoshi, Saitama 351-0198, Japan}
\affiliation{RIKEN Center for Quantum Computing (RQC), Wakoshi, Saitama 351-0198, Japan}

\author{Salvatore Savasta}
\email{ssavasta@unime.it}
\affiliation{Dipartimento di Scienze Matematiche e Informatiche,
Scienze Fisiche e Scienze della Terra, Universit{à} di Messina, I-98166 Messina, Italy}

\author{Carlos S\'anchez Mu\~noz}
\email{carlos.sanchezmunnoz@uam.es}
	\affiliation{Departamento de Física Teórica de la Materia Condensada and Condensed
		Matter Physics Center (IFIMAC), Universidad Autónoma de Madrid, 28049 Madrid,
		Spain}

\begin{abstract}
We show that spontaneous Raman scattering of incident radiation can be observed in cavity-QED systems without external enhancement or coupling to any vibrational degree of freedom. Raman scattering processes can be evidenced as resonances in the emission spectrum, which become clearly visible as the cavity-QED system approaches the ultrastrong coupling regime.
We provide a quantum mechanical description of the effect, and show that ultrastrong light-matter coupling is a necessary condition for the observation of Raman scattering. This effect, and its strong sensitivity to the system parameters, opens new avenues for the characterization of cavity QED setups and the generation of quantum states of light. 
\end{abstract}
\date{\today} \maketitle
%
%
The Raman effect describes the inelastic scattering of radiation by matter, in which scattered photons are produced with a frequency which is either lower (Stokes photons) or larger (anti-Stokes photons) than the frequency of the incident field~\cite{Raman1928,Raman1953,Long2002,Fainstein1997}. In the context of quantum optics and cavity quantum electrodynamics (cQED), this scattering is usually controlled and stimulated via a second resonant drive~\cite{Jamonneau2016,Donarini2019,Vitanov2017,Fleischhauer2005a,Long2018,Guo2019,Wei2008,Shu2009,Kumar2016,Falci2017} or a cavity~\cite{Dimer2007,Sun2018,Hennrich2000,Kuhn2002,Sweeney2014}, 
constituting the basis of many key techniques of coherent control such as
 coherent population trapping~\cite{Jamonneau2016,Donarini2019}, 
 electromagnetic induced transparency~\cite{Fleischhauer2005a,Long2018} or 
 stimulated Raman adiabatic passage~\cite{Wei2008,Shu2009,Kumar2016,Falci2017}.
However, the standard observation of \emph{spontaneous} scattering of Raman photons in the absence of any external stimulation typically arises when the illuminating radiation couples to phonons in a material~\cite{Raman1928,Raman1953,Long2002,Fainstein1997}.  Since this provides a fingerprint of the molecular vibrational modes of the sample, this effect serves as a valuable spectroscopic tool for material characterization~\cite{Orlando2021,Kudelski2008,Pettinger2012,Qian2008,Yang2011,PoornimaParvathi2019}. 
%

Thanks to the plasmonic enhancement of the Raman processes in surface-enhanced Raman spectroscopy~\cite{Langer2020}, single-molecule sensitivity has been achieved~\cite{Kneipp1997,Nie1997}, and state-of-the-art experiments have reached regimes where the quantum nature of the vibrational and electromagnetic modes need to be taken into account~\cite{Zhang2013,Zhu2014,Shalabney2015}. Consequently, several theoretical works have recently  developed  fully quantum mechanical descriptions of Raman scattering~\cite{Schmidt2016,Roelli2016,KamandarDezfouli2017,Dezfouli2019,Hughes2021}, giving rise to the field of molecular optomechanics, where  the interaction between phonons and plasmonic cavity phonons is described as an optomechanical Hamiltonian~\cite{Roelli2020,
Gurlek2021}.

Here, we demonstrate the intriguing possibility of observing spontaneous Raman scattering in cQED systems in which, in stark contrast to molecular optomechanics, there are no vibrational degrees of freedom.  An important feature of the quantum description of Raman scattering is that the underlying process does not conserve the total number of particles: for instance, in a Stokes process, a single laser photon of  given energy will become a single, less energetic photon plus a vibrational excitation---a phonon. The underlying Hamiltonian must therefore not conserve the total number of excitations, which is the case in optomechanical interaction Hamiltonians of the form $\hat V_\mathrm{OM} = g_{\mathrm{OM}} \hat a^\dagger \hat a (\hat b+ \hat b^\dagger ) $ that will arise in molecular optomechanics (with $\aop$ and $\hat b$ annihilation operators of photon and phonon modes, respectively). 

In contrast, let us consider the Hamiltonian describing light-matter interaction between a cavity mode and single dipole, modeled as a two-level system (TLS) with annihilation operator $\sop$. Their coupling is well described by the interaction term of the quantum Rabi model (QRM), $\hat V = g(\hat a + \adop)(\sop + \sdop)$. The counter-rotating terms in $\hat V$ will play no role when the coupling rate is much smaller than the natural frequency of the modes $g \ll \omega_c,\omega_q$, in which case the interaction is well described by the Jaynes-Cummings term $\hat V \approx  g(\hat a \hat\sigma^\dagger + \hat a^\dagger \hat\sigma)$~\cite{scully_book02a,agarwal_book12a}. This interaction term conserves the total number of excitations, and consequently---as we will show---most Raman scattering processes will be forbidden in this regime. The situation changes in the ultrastrong coupling (USC) regime, i.e., the limit where $g$ becomes comparable to $\omega_c$ and $\omega_q$ and the counter-rotating terms $g(\aop\sop + \adop\sdop)$ play an important role in the dynamics~\cite{FriskKockum2019,Forn-Diaz2019}.  Similarly to the optomechanical case, the full QRM that describes the dynamics in the USC does not conserve the total number of excitations, and, as a result, this regime features a wealth of exotic nonlinear processes and applications~\cite{FriskKockum2019,Forn-Diaz2019,Niemczyk2010, Ma2015,Garziano2015,Garziano2016,Stassi2017,Kockum2017,Kockum2017a}.   
We demonstrate that the observation of spontaneous Raman scattering of photons from an incident field is another characteristic process of the USC regime.  In particular, we show unambiguous signatures of these processes in the emission spectra of coherently-driven cQED system in the USC regime~\cite{Salmon2022,Mercurio2022}. This result establish USC-cQED as a novel scenario where Raman Stokes and anti-Stokes photons are produced spontaneously without any vibrational degree of freedom involved. Beyond exact numerical calculations demonstrating the effect, we support these results with predictions from a full quantum description of the process of Raman scattering.

\begin{figure}[t!]
\includegraphics[width=0.5\textwidth]{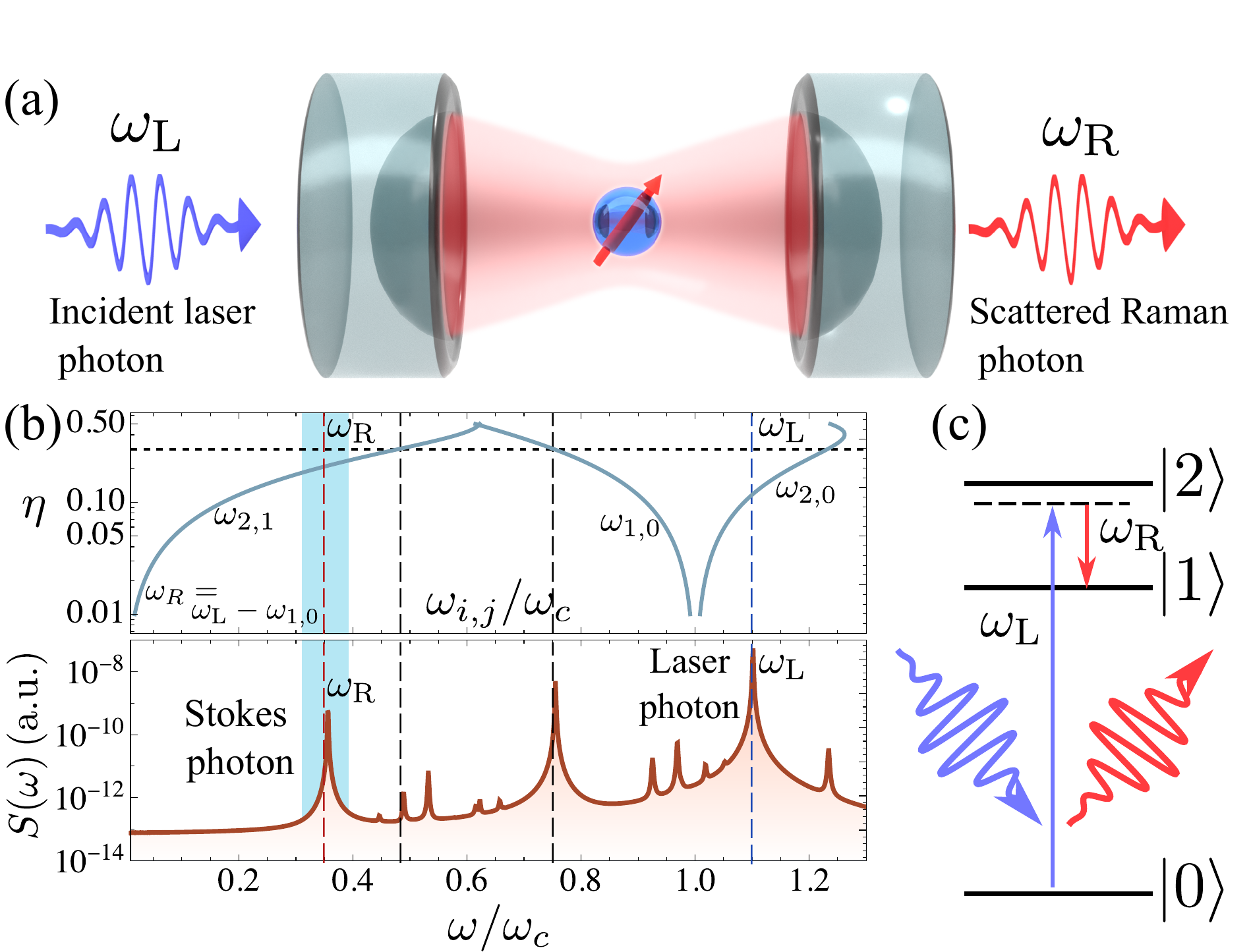}
\caption{a) Scheme of the cavity QED system considered in this work: a quantum emitter interacting with a single cavity mode in the ultrastrong-coupling regime and spontaneously scatters a Raman photon from an incident exciting field. (b) Top: Transition energies between the first two excited eigenstates of the light-matter system and the ground state, versus the normalized coupling parameter $\eta$.  Bottom: Spectrum of emission for $\eta = 0.3$. $\omega_\mathrm L = 1.1\omega_c$. The red line indicates the frequency at which Stokes photons are emitted, originating from the process sketched in (c). }
\label{fig:fig1_setup}
\end{figure}

\emph{Model}---
We consider the cavity QED system sketched in Fig.~\ref{fig:fig1_setup}(a), consisting of a single cavity mode of frequency $\omega_c$, driven by a continuous classical field of frequency $\omega_\mathrm L$, and coupled to a point dipole described in the two-level system (TLS) approximation, with a transition frequency $\omega_q$. Since we will be mostly interested in describing the emission spectrum of such a system, we will resort to the sensor method developed in~\cite{DelValle2012}, adding an ancillary sensor qubit of frequency $\omega_s$ weakly coupled to the cavity, which has been shown to also produce equivalent results to the quantum-regression theorem in the USC limit~\cite{Salmon2021,Salmon2022,Mercurio2022}. The spectrum of emission at frequency $\omega_s$ will thus be proportional to the rate of emission from the sensor qubit. 

The Hamiltonian of this system has the form $\ham = \ham_\text{free}+\ham_I + \ham_\text{drive}$. Setting henceforth $\hbar = 1$, the first term is simply the free Hamiltonian $\ham_\text{free}\equiv \omega_c \adop\aop + \omega_q\sz/2 + \omega_s\sz^s/2$, where $\aop$ is the bosonic annihilation operator of the photon field, and $\hat\sigma_i$ ( $\hat\sigma_i^s$) are standard Pauli operators defined on the TLS (sensor) Hilbert space.
The second term, $\ham_I$, describes light-matter interaction. In order to write it, we define the  polarization operator associated to the TLS and the sensor as $\hat{\bf P} =({\bm\mu}\hat\sigma_p + {\bm \mu}_s \sx^s)\delta(\mathbf r - \mathbf r_0)$, where the TLS operator $\hat\sigma_p$ is given by $\hat\sigma_p \equiv \cos\theta \sx + \sin\theta \sz$, 
and $\bm \mu$ and ${\bm \mu}_s$ are the respective dipole moments of the TLS and sensor. This definition includes the possibility of a TLS with a permanent dipole moment---which breaks the conservation of parity $\hat \Pi = \exp[i\pi(\adop\aop + \sz)]$ in the total system---parametrized  through the angle $\theta$. This parity-breaking term can be induced, i.e., by the flux offset in a flux qubit~\cite{Niemczyk2010} or through strong asymmetries in solid state artificial atoms~\cite{Chestnov2017}.
In the dipole gauge, the interaction Hamiltonian thus takes the form~\cite{Settineri2021}:
\begin{equation}
\ham_I  = i\eta \omega_c (\adop - \aop)\spol + \omega_c\eta_s[i(\adop-\aop)+2\eta\spol]\sx^s,
\label{eq:H_int}
\end{equation}
where $\eta$ and $\eta_s$ are the dimensionless coupling parameters between cavity and TLS and sensor, respectively.
The choice of this gauge ensures a proper gauge invariance under the TLS approximation \cite{DeBernardis2018,DiStefano2019,Salmon2022}.  Finally, the drive Hamiltonian reads $\ham_\text{drive}(t) =\Omega[i(\aop - \adop)-2\eta\sx]\cos(\omega_\mathrm L t)$~\cite{Salmon2022,Mercurio2022}. Further details on the derivation of this Hamiltonian are provided in the Supplemental Material~\footnote{See Supplemental Material for further details on the derivation of the Quantum Rabi Model Hamiltonian in the dipole gauge and on the calculation of time-averaged density matrix form time-dependent Liouvillians using the Floquet method. Includes references~\cite{Settineri2021,Salmon2022,DeBernardis2018,DiStefano2019,Majumdar2011g,Papageorge2012a,Maragkou2013}.}.

Since the setup under consideration is an open quantum system, dissipation must be accounted for by describing the dynamics in terms of a quantum master equation. In the ultrastrong coupling regime, the treatment of dissipation, input-output relationships, correlations, driving, and photo-detection rates requires a proper description of the system-bath interaction in terms of the light-matter eigenstates \cite{Ridolfo2012,Beaudoin2011,DiStefano2018,LeBoite2020}. However, despite the validity of using a dressed master equation with post-trace rotating wave approximation when dealing with USC hybrid systems \cite{Beaudoin2011, Salmon2022}, this approach fails in our case, since the sensor coupling is very low, producing degeneracy and harmonicity. Consequently, following the approach in Refs.~ \cite{Settineri2018, Mercurio2022}, we write a generalized master equation, which is valid at any light-matter coupling strength. In the limit of zero temperature, the master equation reduces to the simple form 
$\dot\dm = -i[\ham,\dm] +\kappa  L_{{\cal \hat X}^+}[\dm]+\gamma  L_{{ \hat \Sigma}^+}[\dm] +\Gamma L_{{ \hat \Sigma_s}^+}[\dm] $, where $L_{\hat O}[\dm]\equiv \hat O \dm \hat O ^\dagger - \{\hat O^\dagger \hat O ,\dm \}/2 $ denotes the standard Lindblad terms, and the decay operators 
are given by
 $\hat {\cal X}^+ = \sum_{j=1} \sum_{k >j}
\langle j| [i  (\hat a - \hat a^{\dag}) - 2   \eta  \hat \sigma_x ]|  k \rangle | j \rangle \langle k |$, 
$\hat {\Sigma}_{(s)}^+ = i \sum_{j=1} \sum_{k >j}\langle j|  \sx^{(s)} |  k \rangle | j \rangle \langle k | \omega_{kj}/\omega_q$, 
describing the decay from the cavity, the TLS and the sensor with decay rates $\kappa$, $\gamma$, and $\Gamma$, respectively. 
 
Due to the presence of counter-rotating terms in \eqref{eq:H_int}, applying the standard unitary transformation  $e^{i\omega_\mathrm L a^\dagger a}$ to the laser frame will not eliminate the time dependence in the Hamiltonian. Such an oscillating Hamiltonian will yield, in the long time limit, a time-dependent density matrix oscillating around an average steady state $\hat\rho_\mathrm{ss} +\delta\hat\rho(t)$. 
Here, we consider the limit of a very small drive $\Omega\ll \eta \omega_c$, so that these oscillations become negligible, and thus we focus on the average steady state $\hat \rho_\text{ss}$~\cite{Note1}. We consider that the stationary rate of emission from the sensor is proportional to the spectrum of emission at the sensor's frequency, i.e. $S(\omega_s) \propto \text{Tr}[\hat \rho_{\mathrm ss}\hat\Sigma^-_s \hat\Sigma_s^+]$, with the sensor's decay rate $\Gamma$ corresponding to the filter linewidth. 

 \begin{figure*}[t!]
\includegraphics[width=0.98\textwidth]{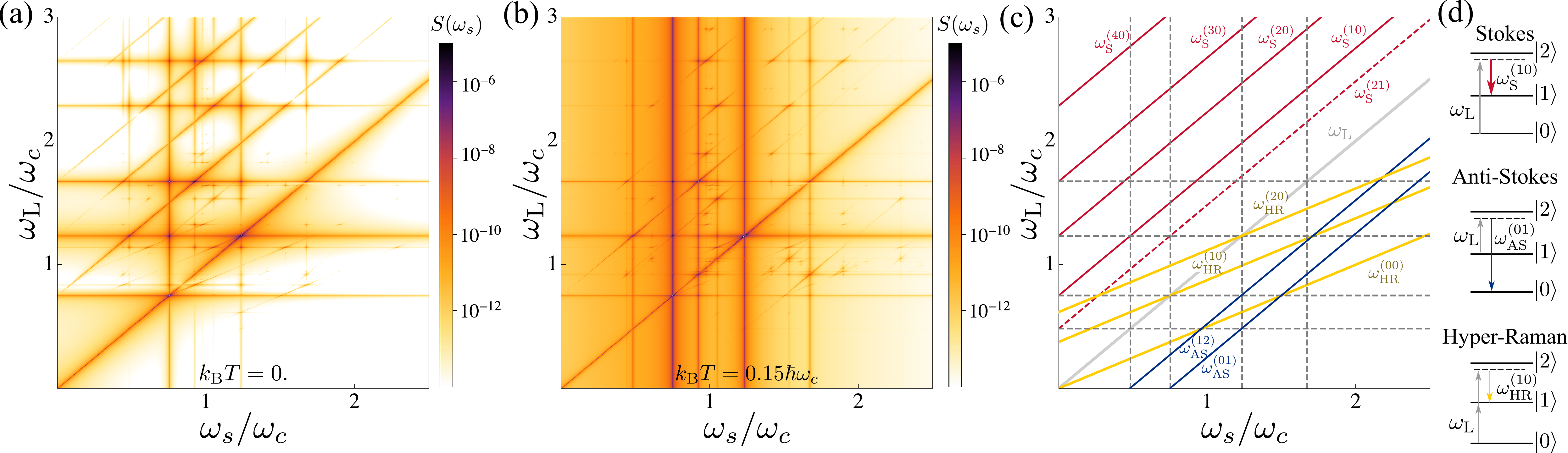}
\caption{Emission of Raman photons evidenced through excitation-emission spectra. (a) Excitation-emission spectrum at zero temperature. Raman photons are revealed as diagonal lines (i.e., peaks with frequencies that depend on the laser frequency). (b) Excitation-emission spectrum at non-zero temperature $k_\mathrm{B}T=0.15\hbar\omega_c$. Anti-stokes peaks, which rely on the non-zero population of exited states, are enhanced in this case. (c) Identification of some of the processes seen in panels (a) and (b), which include Stokes, anti-Stokes and hyper-Raman photons. Parameters: $\omega_q = \omega_c$, $\theta = \pi / 6$, $\eta = 0.3$, $\eta_s = 10^{-5}$, $\Omega = 5 \times 10^{-3} \omega_c$, $\kappa = \gamma = \Gamma = 10^{-3} \omega_c$.}
\label{fig:fig2_2d}
\end{figure*}

\emph{Scattering of Raman photons---}
Figure~\ref{fig:fig1_setup}(b) depicts an example of the emission spectrum. Here and in the following we fix, unless stated otherwise, $\omega_q = \omega_c$, $\theta = \pi / 6$, $\eta = 0.3$, $\eta_s = 10^{-5}$, $\Omega = 5 \times 10^{-3} \omega_c$, $\kappa = \gamma = \Gamma = 10^{-3} \omega_c$, and $\omega_\mathrm L = 1.1\omega_c$. At first glance, one can observe the presence of a resonance peak at the cavity frequency, and further peaks that match transition energies between the light-matter eigenstates, displayed in the top panel of Fig.~\ref{fig:fig1_setup}(b). In addition to these, one can observe an additional peak that corresponds to the spontaneous scattering of a Stokes photon. The corresponding Raman process that gives rise to this peak is sketched in Fig.~\ref{fig:fig1_setup}(c). Via a second-order process, an input laser photon of frequency $\omega_\mathrm L$ is converted into a lower-energy Raman photon of energy $\omega_\mathrm R$ and a light-matter excitation of energy $\omega_{1}$. Since energy must be conserved in the whole process, the energy of the Stokes photon is expected to be $\omega_{\mathrm R} = \omega_\mathrm L - \omega_{1}$, and thus it depends linearly with the laser excitation. 

In order to understand the emergence of Raman peaks and its dependence on system parameters such as $g$ or $\theta$, we develop here a full quantum description of the Raman scattering process. To do this, we consider that the cavity is coupled to a broad quasi-continuum of modes with  $\hat H_b =  \sum_q \omega_q \hat b_q^\dag \hat b_q$, which will contain the incident radiation field and the scattered Raman photons. The total system Hamiltonian is 
$\hat H_\mathrm{total} = \hat H_R + \hat H_b + \hat V_b$, 
where $\ham_R$ is the quantum Rabi Hamiltonian ($\ham$ above, without the sensor and the drive terms) and $\hat V_{b} =  \sum_q g_q (\hat b_q + \hat b_q^\dag) (\aop + \adop)$. In the following, we consider 
$\hat H_R + \hat H_b$ as the unperturbed, bare Hamiltonian, and we express $\hat H_R$  in diagonal form as $\hat H_R =  \sum_j \omega_j\,  | j \rangle \langle j|$, 
where we chose the labeling of the states such that $\omega_k > \omega_j$ for $k >j$.
The Raman scattering process can be described by second-order perturbation theory under the constant perturbation $\hat V_{b}$. 
Let us consider an initial state $|I_i \rangle = |i, \nl, \nr \rangle$, where the first entry labels the eigenstates of $\hat H_R$, $n_\mathrm L$ labels the photon number in the input mode---the laser drive---with frequency $\omega_\mathrm L$, and $n_\mathrm R$ indicates the photon number in the output mode of frequency $\omega_\mathrm R$, where Raman photons are being emitted. We are considering here only the two modes involved in the scattering process; all the other modes of the quasi-continuum are assumed to be in the zero-photon states throughout the process. The energy of the initial state is $\omega_{I,i} = \omega_i  +  \omega_\mathrm L n_\mathrm L +  \omega_\mathrm R n_\mathrm R $.
Then, we consider a final state  $| F_f \rangle =  |f, \nl-1, \nr +1 \rangle$, with energy $\omega_{F,f} =  \omega_f  +  \oml(\nl-1) + \omr (\nr +1) $. Energy conservation implies $\omega_{F,f} = \omega_{I,i}$, and, therefore, for a particular choice of initial and final states $i$ and $f$, the energy of the corresponding Raman photons is
\begin{equation}
\omr = \omr^{(f,i)}\equiv \oml - (\omega_f - \omega_i).
\end{equation}
 $| F_f \rangle$  is connected to the initial state $| I_i \rangle$ by a second-order process involving an intermediate virtual state. It is possible to identify two kinds of intermediate states, $|T_{1}\rangle$ and  $|T_{2}\rangle$, describing respectively the process (i) where a photon is first absorbed from the input state: $|T_1 \rangle = |j, \nl -1, \nr \rangle$, with energy $\omega_{T_1} = \omega_j  +  \oml (\nl-1) + \omr \nr$ ; and the process (ii) where a photon is first emitted into the output mode:  $|T_2 \rangle = |j, \nl, \nr+1 \rangle$, with energy  $\omega_{T_2} = \omega_j  +\oml \nl +\omr (\nr+1)$.

The rate of the process $|I_i\rangle \rightarrow |F_f\rangle$ given by the Fermi golden rule, for a given $\oml$, $i$ and $f$, is
\begin{equation}
W_{f,i}(\oml,\omr) = \frac{2\pi}{\hbar} \gr^2 \gl^2 \nl (\nr+1)|M_{f,i}|^2 \delta(\omr-\omr^{(f,i)}),
\label{eq:Wik}
\end{equation}
where $\omega_{f,i}=\omega_f - \omega_i$ and 
\begin{equation}
M_{f,i}(\oml,\omr) = \sum_j \left(\frac{X_{f,j}X_{j,i}}{\omega_{T_1}- \omega_{I_i}} + \frac{X_{f,j}X_{j,i}}{\omega_{T_2}- \omega_{I_i}}\right),
\end{equation}
with $X_{f,j}\equiv \langle f |\aop + \adop |j\rangle$. Notice that $\omega_{T_1} - \omega_{I,i} = \omega_{j,i}  -\oml$, and $\omega_{T_2} - \omega_{I,i} = \omega_{j,i} +\omr^{(f,i)}$. The total scattering rate for the process is obtained by summing over all possible initial and final states, which will be constrained by the energy-conservation condition in \eqref{eq:Wik}, giving
\begin{equation}
W(\oml,\omr) = \sum_{f,i}W_{f,i}(\oml,\omr) \rho^\mathrm{ss}_{i}(1-\rho^\mathrm{ss}_f),
\label{eq:wrl}
\end{equation}
where $\rho^\mathrm{ss}_k$ is the steady-state occupation probability of the eigenstate $|k\rangle$ of $\ham_R$. For a system at very low temperatures and low driving which is mostly in the ground state, so that $\rho^\mathrm{ss}_0 \approx 1$, we obtain $W(\oml,\omr) = \sum_{f}W_{f,0}(\oml,\omr)$. Therefore, if Raman spectroscopy is performed by probing cQED systems that are close to the ground state, only the family of Raman processes that start from $|0\rangle$ are expected to be observed.

\begin{figure}[b!]
\includegraphics[width=0.49\textwidth]{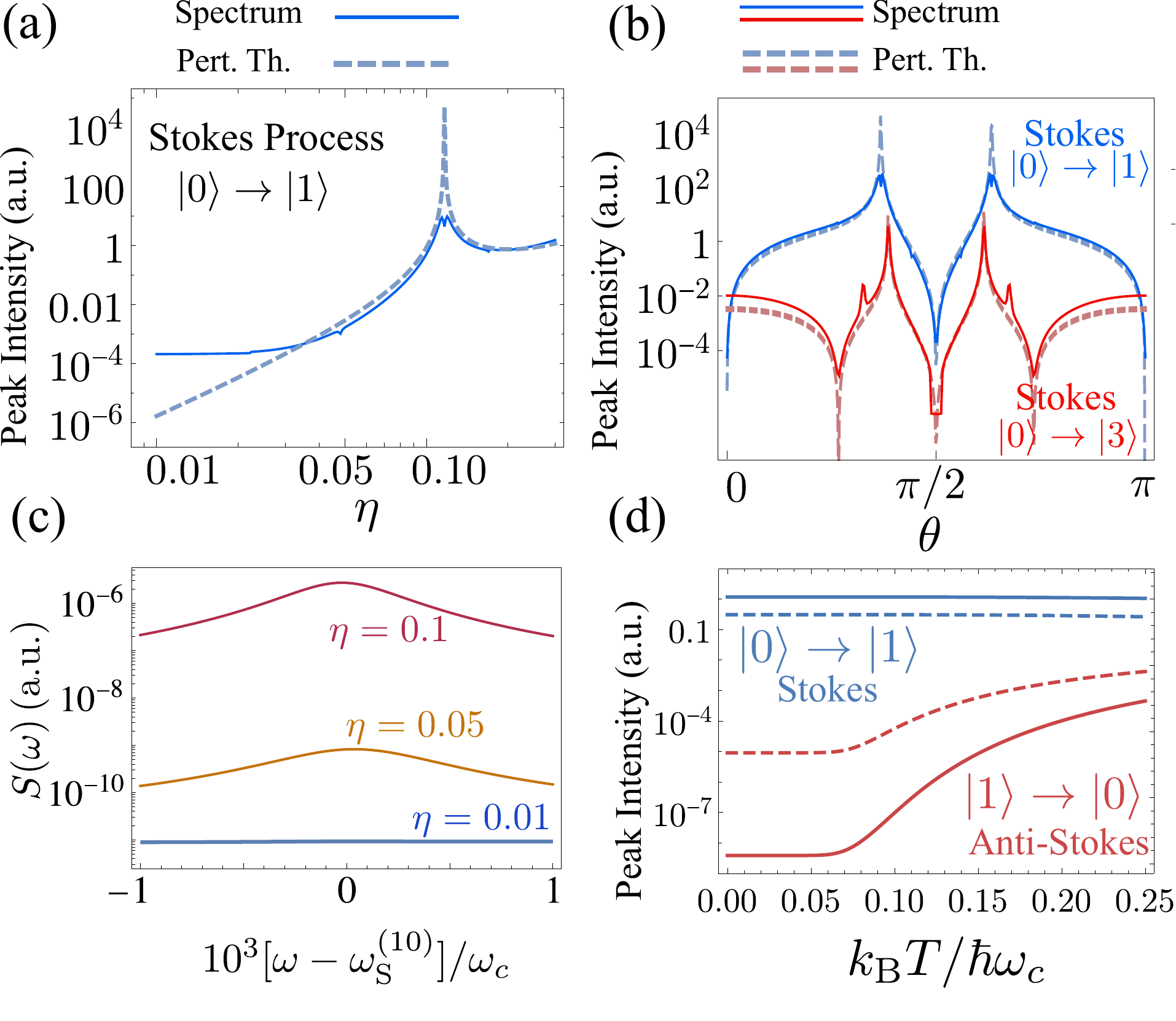}
\caption{(a) Intensity of the Raman peak $\omr^{(10)}$ and corresponding scattering rate from Fermi's Golden rule.  $\theta = \pi/6$. (b)  Raman intensity versus $\theta$ for two types Stokes processes. (c) Spectrum around Raman resonance $\omega_\text{S}^{(10)}$ for increasing $\eta$. (d) Raman intensity versus temperature for a Stokes process and an anti-Stokes process.}
\label{fig:fig3}
\end{figure}

\emph{Visibility of Raman processes---.}
The quantum scattering process outlined above manifests as resonances in the spectrum of emission, centered at the frequencies $\omr^{(f,i)}$. These peaks have a characteristic feature that distinguishes them from peaks arising from standard radiative transitions: their central frequency $\omr^{(f,i)}$ depends linearly on the laser frequency $\omega_\mathrm L$. The linear dependence between the $\omr^{(f,i)}$ and $\omega_\mathrm L$ manifests as resonance peaks that follow straight lines in the excitation-emission spectrum, i.e., the spectra of emission for different driving frequencies.  This feature represents an unambiguous proof that these peaks correspond to the inelastic, spontaneous scattering of laser photons through the quantum process outlined above.   Numerical calculations of the resulting excitation-emission spectra for a cavity-QED system are shown in Fig.~\ref{fig:fig2_2d}(a-b) at two different temperatures. The Raman peaks that are most clearly identified are labeled in Fig.~\ref{fig:fig2_2d}(c) . At low temperatures the most visible peaks are Stokes processes that start at the ground state of the light-matter system and end at some excited state $|f\rangle$ (we label these Stokes processes as $\omega_\mathrm S^{(f0)}$). In agreement to what is expected from ~\eqref{eq:wrl}, at small temperatures, Stokes processes that start in an excited state are hardly visible or not visible at all; in Fig.~\ref{fig:fig2_2d}(c)  we highlight the process $\omega_\mathrm S^{(21)}$---starting in $|1\rangle$ and finishing in $|2\rangle$---which is the one that can be recognized in certain regions of the spectra shown. Likewise, the emission of anti-Stokes photons with frequencies larger than the drive frequency is only clearly visible at finite temperatures: these processes require the energy of the final state of the cQED system to be lower than the initial one, and therefore, the initial state needs to be an excited state with a non-negligible stationary occupation probability. These calculations also show that higher-order, hyper-Raman processes are faintly visible as well in the excitation emission spectra. These processes scatter two incident laser photons into a Raman photon, and therefore conservation of energy establishes that the frequency of the hyper-Raman photons must be $\omega_\mathrm R^{(fi)}=2\omega_\mathrm L-(\omega_f-\omega_i)$. Such processes are then identified in the excitation-emission spectra as straight lines with twice the slope of standard Raman processes.

All the features just outlined are clearer when one approaches the ultra-strong coupling regime of light-matter interaction, $\eta \sim 0.1$, so that the matrix elements $X_{k,j}$ in \eqref{eq:Wik} acquire sizable values.
Indeed, Fig.~\ref{fig:fig3}(a) shows the calculation of the scattering rate $W$ versus $\eta$ computed through \eqref{eq:wrl} for the Stokes process $|0\rangle \rightarrow |1\rangle$, which is the most visible one in a system close to the ground state, compared to the intensity of the corresponding Raman peak computed in the excitation emission spectra. Beyond the good agreement between both results, which supports our description of the underlying quantum process, we highlight the exponential increase of the intensity of the peak with $\eta$.
Way bellow the USC regime, the small values of the scattering rate would make observing Raman processes in cavity QED systems very challenging, as shown in Fig.~\ref{fig:fig3}(c), where for $\eta=0.01$ the first Stokes peak is extremely hard to notice.

It is illustrative to consider the possibility of Raman processes in a cQED system with $\eta\ll 1$, therefore well described by a Jaynes-Cummings Hamiltonian. The eigenstates of this system are organized in doublets $|j_\pm\rangle$ that are also eigenstates of the total number of excitations $\hat N = \adop \aop +\sdop\sop $, i.e. $\hat N |j_\pm\rangle = j|j_\pm\rangle$. This means that the only Raman processes allowed are those that conserve the total number of excitations, i.e., those whose initial and final states are within the same doublet. Since processes that start and end in the ground state yield $\omega_\mathrm{R} = \omega_\mathrm{L}$ and therefore do not produce energy-shifted photons, the observation of the most relevant Raman processes that involve the ground state is not possible in Jaynes Cummings system. Peaks that may be observed in this limit, such as $\omega_\mathrm{S}^{21}$ and $\omega_\mathrm{AS}^{21}$, are only vaguely visible even in the ultrastrong-coupling regime, as can be seen in Fig.~\ref{fig:fig2_2d}(a,b), and would require a stationary population of excited states, whose origin can imply extra sources of dephasing. We therefore conclude that emission of Raman photons from coherently driven cavity QED systems is essentially a characteristic effect of the USC coupling limit.

The presence or absence of certain Raman peaks also provides information about microscopic parameters, such as the static dipole moment parametrized by $\theta$.
Each peak exhibits a characteristic dependence on $\theta$, as we illustrate in Fig.~\ref{fig:fig3}(b) with two particular examples, the Stokes peaks  $\omega_\mathrm{S}^{(10)}$ and $\omega_\mathrm{S}^{(30)}$, showing that this dependence is well captured by our quantum description of the process based on perturbation theory. This example highlights that, in some cases---such as for $\omega_\mathrm{R}^{(10)}$---the breaking of parity symmetry ($\theta \neq 0$) is necessary to observe the corresponding Raman peak. For $\theta=0$, eigenstates of the QRM are also parity eigenstates, and thus only Raman processes that conserve parity, such as $|0\rangle\rightarrow|3\rangle$, will have a non-zero scattering rate. 

Finally, Fig.~\ref{fig:fig3}(d) shows that our quantum model provides a good qualitative prediction for the different dependence of Stokes and anti-Stokes peaks on temperature, showing that the intensity Stokes peaks is just slightly reduced, while the intensity of anti-Stokes peaks can be increased by orders of magnitude, explained by the corresponding increase of the stationary population of excited states. 

\emph{Conclusions---.} We have demonstrated that spontaneous scattering of Raman photons from cQED systems can be visible in the USC regime without involving any vibrational degree of freedom. This result introduces news fingerprints of strong light-matter interaction that will allow to leverage the potential of Raman spectroscopy for system characterization in the field of cQED. The study of quantum correlations in Raman photons can also offer new  routes for the generation of non-classical light~\cite{Faraon2008,Ridolfo2012,Chang2014a,Muller2015a,Hamsen2017}.

\begin{acknowledgments}
\emph{Acknowledgments}---The authors thank D. Martin-Cano and S. Hughes
 for useful feedback.   C.S.M. acknowledges that the project that
gave rise to these results received the support of a fellowship from “la Caixa” Foundation (ID 100010434) and from the
European Union’s Horizon 2020 research and innovation programme under the Marie Skłodowska-Curie Grant Agreement
No. 847648, with fellowship code LCF/BQ/PI20/11760026, and financial support from the Proyecto Sinérgico CAM 2020
Y2020/TCS-6545 (NanoQuCo-CM). 
F.N. is supported in part by: Nippon Telegraph and Telephone Corporation (NTT) Research, the
Japan Science and Technology Agency (JST) [via the Quantum Leap Flagship Program (Q-LEAP) program,
and the Moonshot R\&D Grant Number JPMJMS2061],
the Japan Society for the Promotion of Science (JSPS)
[via the Grants-in- Aid for Scientific Research (KAKENHI) Grant No. JP20H00134], the Army Research
Office (ARO) (Grant No. W911NF- 18-1-0358), the
Asian Office of Aerospace Research and Development
(AOARD) (via Grant No. FA2386-20-1- 4069), and
the Foundational Questions Institute Fund (FQXi) via
Grant No. FQXi-IAF19-06. 
S.S. acknowledges the Army Research Office (ARO) (Grant No. W911NF1910065).
\end{acknowledgments}

\bibliography{library,books}

\begin{thebibliography}{70}%
\makeatletter
\providecommand \@ifxundefined [1]{%
 \@ifx{#1\undefined}
}%
\providecommand \@ifnum [1]{%
 \ifnum #1\expandafter \@firstoftwo
 \else \expandafter \@secondoftwo
 \fi
}%
\providecommand \@ifx [1]{%
 \ifx #1\expandafter \@firstoftwo
 \else \expandafter \@secondoftwo
 \fi
}%
\providecommand \natexlab [1]{#1}%
\providecommand \enquote  [1]{``#1''}%
\providecommand \bibnamefont  [1]{#1}%
\providecommand \bibfnamefont [1]{#1}%
\providecommand \citenamefont [1]{#1}%
\providecommand \href@noop [0]{\@secondoftwo}%
\providecommand \href [0]{\begingroup \@sanitize@url \@href}%
\providecommand \@href[1]{\@@startlink{#1}\@@href}%
\providecommand \@@href[1]{\endgroup#1\@@endlink}%
\providecommand \@sanitize@url [0]{\catcode `\\12\catcode `\$12\catcode
  `\&12\catcode `\#12\catcode `\^12\catcode `\_12\catcode `\%12\relax}%
\providecommand \@@startlink[1]{}%
\providecommand \@@endlink[0]{}%
\providecommand \url  [0]{\begingroup\@sanitize@url \@url }%
\providecommand \@url [1]{\endgroup\@href {#1}{\urlprefix }}%
\providecommand \urlprefix  [0]{URL }%
\providecommand \Eprint [0]{\href }%
\providecommand \doibase [0]{https://doi.org/}%
\providecommand \selectlanguage [0]{\@gobble}%
\providecommand \bibinfo  [0]{\@secondoftwo}%
\providecommand \bibfield  [0]{\@secondoftwo}%
\providecommand \translation [1]{[#1]}%
\providecommand \BibitemOpen [0]{}%
\providecommand \bibitemStop [0]{}%
\providecommand \bibitemNoStop [0]{.\EOS\space}%
\providecommand \EOS [0]{\spacefactor3000\relax}%
\providecommand \BibitemShut  [1]{\csname bibitem#1\endcsname}%
\let\auto@bib@innerbib\@empty
\bibitem [{\citenamefont {Raman}\ and\ \citenamefont
  {Krishnan}(1928)}]{Raman1928}%
  \BibitemOpen
  \bibfield  {author} {\bibinfo {author} {\bibfnamefont {C.~V.}\ \bibnamefont
  {Raman}}\ and\ \bibinfo {author} {\bibfnamefont {K.~S.}\ \bibnamefont
  {Krishnan}},\ }\bibfield  {title} {\bibinfo {title} {{A New Type of Secondary
  Radiation}},\ }\href {https://doi.org/10.1038/121501c0} {\bibfield  {journal}
  {\bibinfo  {journal} {Nature}\ }\textbf {\bibinfo {volume} {121}},\ \bibinfo
  {pages} {501} (\bibinfo {year} {1928})}\BibitemShut {NoStop}%
\bibitem [{\citenamefont {Raman}(1953)}]{Raman1953}%
  \BibitemOpen
  \bibfield  {author} {\bibinfo {author} {\bibfnamefont {C.~V.}\ \bibnamefont
  {Raman}},\ }\bibfield  {title} {\bibinfo {title} {{A new radiation}},\ }\href
  {https://doi.org/10.1007/BF03052651} {\bibfield  {journal} {\bibinfo
  {journal} {Proceedings of the Indian Academy of Sciences - Section A}\
  }\textbf {\bibinfo {volume} {37}},\ \bibinfo {pages} {333} (\bibinfo {year}
  {1953})}\BibitemShut {NoStop}%
\bibitem [{\citenamefont {Long}(2002)}]{Long2002}%
  \BibitemOpen
  \bibfield  {author} {\bibinfo {author} {\bibfnamefont {D.~A.}\ \bibnamefont
  {Long}},\ }\href {https://doi.org/10.1002/0470845767} {\emph {\bibinfo
  {title} {{The Raman Effect: A Unified Treatment of the Theory of Raman
  Scattering by Molecules}}}}\ (\bibinfo  {publisher} {John Wiley \& Sons,
  Ltd},\ \bibinfo {address} {Chichester, UK},\ \bibinfo {year}
  {2002})\BibitemShut {NoStop}%
\bibitem [{\citenamefont {Fainstein}\ \emph {et~al.}(1997)\citenamefont
  {Fainstein}, \citenamefont {Jusserand},\ and\ \citenamefont
  {Thierry-Mieg}}]{Fainstein1997}%
  \BibitemOpen
  \bibfield  {author} {\bibinfo {author} {\bibfnamefont {A.}~\bibnamefont
  {Fainstein}}, \bibinfo {author} {\bibfnamefont {B.}~\bibnamefont
  {Jusserand}},\ and\ \bibinfo {author} {\bibfnamefont {V.}~\bibnamefont
  {Thierry-Mieg}},\ }\bibfield  {title} {\bibinfo {title} {{Cavity-Polariton
  Mediated Resonant Raman Scattering}},\ }\href
  {https://doi.org/10.1103/PhysRevLett.78.1576} {\bibfield  {journal} {\bibinfo
   {journal} {Physical Review Letters}\ }\textbf {\bibinfo {volume} {78}},\
  \bibinfo {pages} {1576} (\bibinfo {year} {1997})}\BibitemShut {NoStop}%
\bibitem [{\citenamefont {Jamonneau}\ \emph {et~al.}(2016)\citenamefont
  {Jamonneau}, \citenamefont {H{\'{e}}tet}, \citenamefont {Dr{\'{e}}au},
  \citenamefont {Roch},\ and\ \citenamefont {Jacques}}]{Jamonneau2016}%
  \BibitemOpen
  \bibfield  {author} {\bibinfo {author} {\bibfnamefont {P.}~\bibnamefont
  {Jamonneau}}, \bibinfo {author} {\bibfnamefont {G.}~\bibnamefont
  {H{\'{e}}tet}}, \bibinfo {author} {\bibfnamefont {A.}~\bibnamefont
  {Dr{\'{e}}au}}, \bibinfo {author} {\bibfnamefont {J.-F.}\ \bibnamefont
  {Roch}},\ and\ \bibinfo {author} {\bibfnamefont {V.}~\bibnamefont
  {Jacques}},\ }\bibfield  {title} {\bibinfo {title} {{Coherent Population
  Trapping of a Single Nuclear Spin Under Ambient Conditions}},\ }\href
  {https://doi.org/10.1103/PhysRevLett.116.043603} {\bibfield  {journal}
  {\bibinfo  {journal} {Physical Review Letters}\ }\textbf {\bibinfo {volume}
  {116}},\ \bibinfo {pages} {043603} (\bibinfo {year} {2016})}\BibitemShut
  {NoStop}%
\bibitem [{\citenamefont {Donarini}\ \emph {et~al.}(2019)\citenamefont
  {Donarini}, \citenamefont {Niklas}, \citenamefont {Schafberger},
  \citenamefont {Paradiso}, \citenamefont {Strunk},\ and\ \citenamefont
  {Grifoni}}]{Donarini2019}%
  \BibitemOpen
  \bibfield  {author} {\bibinfo {author} {\bibfnamefont {A.}~\bibnamefont
  {Donarini}}, \bibinfo {author} {\bibfnamefont {M.}~\bibnamefont {Niklas}},
  \bibinfo {author} {\bibfnamefont {M.}~\bibnamefont {Schafberger}}, \bibinfo
  {author} {\bibfnamefont {N.}~\bibnamefont {Paradiso}}, \bibinfo {author}
  {\bibfnamefont {C.}~\bibnamefont {Strunk}},\ and\ \bibinfo {author}
  {\bibfnamefont {M.}~\bibnamefont {Grifoni}},\ }\bibfield  {title} {\bibinfo
  {title} {{Coherent population trapping by dark state formation in a carbon
  nanotube quantum dot}},\ }\href {https://doi.org/10.1038/s41467-018-08112-x}
  {\bibfield  {journal} {\bibinfo  {journal} {Nature Communications}\ }\textbf
  {\bibinfo {volume} {10}},\ \bibinfo {pages} {381} (\bibinfo {year}
  {2019})}\BibitemShut {NoStop}%
\bibitem [{\citenamefont {Vitanov}\ \emph {et~al.}(2017)\citenamefont
  {Vitanov}, \citenamefont {Rangelov}, \citenamefont {Shore},\ and\
  \citenamefont {Bergmann}}]{Vitanov2017}%
  \BibitemOpen
  \bibfield  {author} {\bibinfo {author} {\bibfnamefont {N.~V.}\ \bibnamefont
  {Vitanov}}, \bibinfo {author} {\bibfnamefont {A.~A.}\ \bibnamefont
  {Rangelov}}, \bibinfo {author} {\bibfnamefont {B.~W.}\ \bibnamefont
  {Shore}},\ and\ \bibinfo {author} {\bibfnamefont {K.}~\bibnamefont
  {Bergmann}},\ }\bibfield  {title} {\bibinfo {title} {{Stimulated Raman
  adiabatic passage in physics, chemistry, and beyond}},\ }\href
  {https://doi.org/10.1103/RevModPhys.89.015006} {\bibfield  {journal}
  {\bibinfo  {journal} {Reviews of Modern Physics}\ }\textbf {\bibinfo {volume}
  {89}},\ \bibinfo {pages} {015006} (\bibinfo {year} {2017})}\BibitemShut
  {NoStop}%
\bibitem [{\citenamefont {Fleischhauer}\ \emph {et~al.}(2005)\citenamefont
  {Fleischhauer}, \citenamefont {Imamoglu},\ and\ \citenamefont
  {Marangos}}]{Fleischhauer2005a}%
  \BibitemOpen
  \bibfield  {author} {\bibinfo {author} {\bibfnamefont {M.}~\bibnamefont
  {Fleischhauer}}, \bibinfo {author} {\bibfnamefont {A.}~\bibnamefont
  {Imamoglu}},\ and\ \bibinfo {author} {\bibfnamefont {P.~J.}\ \bibnamefont
  {Marangos}},\ }\bibfield  {title} {\bibinfo {title} {{Electromagnetically
  induced transparency}},\ }\href
  {https://doi.org/10.1103/REVMODPHYS.77.633/FIGURES/23/MEDIUM} {\bibfield
  {journal} {\bibinfo  {journal} {Reviews of Modern Physics}\ }\textbf
  {\bibinfo {volume} {77}},\ \bibinfo {pages} {633} (\bibinfo {year}
  {2005})}\BibitemShut {NoStop}%
\bibitem [{\citenamefont {Long}\ \emph {et~al.}(2018)\citenamefont {Long},
  \citenamefont {Ku}, \citenamefont {Wu}, \citenamefont {Gu}, \citenamefont
  {Lake}, \citenamefont {Bal}, \citenamefont {Liu},\ and\ \citenamefont
  {Pappas}}]{Long2018}%
  \BibitemOpen
  \bibfield  {author} {\bibinfo {author} {\bibfnamefont {J.}~\bibnamefont
  {Long}}, \bibinfo {author} {\bibfnamefont {H.~S.}\ \bibnamefont {Ku}},
  \bibinfo {author} {\bibfnamefont {X.}~\bibnamefont {Wu}}, \bibinfo {author}
  {\bibfnamefont {X.}~\bibnamefont {Gu}}, \bibinfo {author} {\bibfnamefont
  {R.~E.}\ \bibnamefont {Lake}}, \bibinfo {author} {\bibfnamefont
  {M.}~\bibnamefont {Bal}}, \bibinfo {author} {\bibfnamefont {Y.~X.}\
  \bibnamefont {Liu}},\ and\ \bibinfo {author} {\bibfnamefont {D.~P.}\
  \bibnamefont {Pappas}},\ }\bibfield  {title} {\bibinfo {title}
  {{Electromagnetically Induced Transparency in Circuit Quantum Electrodynamics
  with Nested Polariton States}},\ }\href
  {https://doi.org/10.1103/PHYSREVLETT.120.083602/FIGURES/5/MEDIUM} {\bibfield
  {journal} {\bibinfo  {journal} {Physical Review Letters}\ }\textbf {\bibinfo
  {volume} {120}},\ \bibinfo {pages} {083602} (\bibinfo {year}
  {2018})}\BibitemShut {NoStop}%
\bibitem [{\citenamefont {Guo}\ \emph {et~al.}(2019)\citenamefont {Guo},
  \citenamefont {Shu}, \citenamefont {Dong},\ and\ \citenamefont
  {Nori}}]{Guo2019}%
  \BibitemOpen
  \bibfield  {author} {\bibinfo {author} {\bibfnamefont {Y.}~\bibnamefont
  {Guo}}, \bibinfo {author} {\bibfnamefont {C.-C.}\ \bibnamefont {Shu}},
  \bibinfo {author} {\bibfnamefont {D.}~\bibnamefont {Dong}},\ and\ \bibinfo
  {author} {\bibfnamefont {F.}~\bibnamefont {Nori}},\ }\bibfield  {title}
  {\bibinfo {title} {{Vanishing and Revival of Resonance Raman Scattering}},\
  }\href {https://doi.org/10.1103/PhysRevLett.123.223202} {\bibfield  {journal}
  {\bibinfo  {journal} {Physical Review Letters}\ }\textbf {\bibinfo {volume}
  {123}},\ \bibinfo {pages} {223202} (\bibinfo {year} {2019})}\BibitemShut
  {NoStop}%
\bibitem [{\citenamefont {Wei}\ \emph {et~al.}(2008)\citenamefont {Wei},
  \citenamefont {Johansson}, \citenamefont {Cen}, \citenamefont {Ashhab},\ and\
  \citenamefont {Nori}}]{Wei2008}%
  \BibitemOpen
  \bibfield  {author} {\bibinfo {author} {\bibfnamefont {L.~F.}\ \bibnamefont
  {Wei}}, \bibinfo {author} {\bibfnamefont {J.~R.}\ \bibnamefont {Johansson}},
  \bibinfo {author} {\bibfnamefont {L.~X.}\ \bibnamefont {Cen}}, \bibinfo
  {author} {\bibfnamefont {S.}~\bibnamefont {Ashhab}},\ and\ \bibinfo {author}
  {\bibfnamefont {F.}~\bibnamefont {Nori}},\ }\bibfield  {title} {\bibinfo
  {title} {{Controllable Coherent Population Transfers in Superconducting
  Qubits for Quantum Computing}},\ }\href
  {https://doi.org/10.1103/PhysRevLett.100.113601} {\bibfield  {journal}
  {\bibinfo  {journal} {Physical Review Letters}\ }\textbf {\bibinfo {volume}
  {100}},\ \bibinfo {pages} {113601} (\bibinfo {year} {2008})}\BibitemShut
  {NoStop}%
\bibitem [{\citenamefont {Shu}\ \emph {et~al.}(2009)\citenamefont {Shu},
  \citenamefont {Yu}, \citenamefont {Yuan}, \citenamefont {Hu}, \citenamefont
  {Yang},\ and\ \citenamefont {Cong}}]{Shu2009}%
  \BibitemOpen
  \bibfield  {author} {\bibinfo {author} {\bibfnamefont {C.-C.}\ \bibnamefont
  {Shu}}, \bibinfo {author} {\bibfnamefont {J.}~\bibnamefont {Yu}}, \bibinfo
  {author} {\bibfnamefont {K.-J.}\ \bibnamefont {Yuan}}, \bibinfo {author}
  {\bibfnamefont {W.-H.}\ \bibnamefont {Hu}}, \bibinfo {author} {\bibfnamefont
  {J.}~\bibnamefont {Yang}},\ and\ \bibinfo {author} {\bibfnamefont {S.-L.}\
  \bibnamefont {Cong}},\ }\bibfield  {title} {\bibinfo {title} {{Stimulated
  Raman adiabatic passage in molecular electronic states}},\ }\href
  {https://doi.org/10.1103/PhysRevA.79.023418} {\bibfield  {journal} {\bibinfo
  {journal} {Physical Review A}\ }\textbf {\bibinfo {volume} {79}},\ \bibinfo
  {pages} {023418} (\bibinfo {year} {2009})}\BibitemShut {NoStop}%
\bibitem [{\citenamefont {Kumar}\ \emph {et~al.}(2016)\citenamefont {Kumar},
  \citenamefont {Veps{\"{a}}l{\"{a}}inen}, \citenamefont {Danilin},\ and\
  \citenamefont {Paraoanu}}]{Kumar2016}%
  \BibitemOpen
  \bibfield  {author} {\bibinfo {author} {\bibfnamefont {K.~S.}\ \bibnamefont
  {Kumar}}, \bibinfo {author} {\bibfnamefont {A.}~\bibnamefont
  {Veps{\"{a}}l{\"{a}}inen}}, \bibinfo {author} {\bibfnamefont
  {S.}~\bibnamefont {Danilin}},\ and\ \bibinfo {author} {\bibfnamefont {G.~S.}\
  \bibnamefont {Paraoanu}},\ }\bibfield  {title} {\bibinfo {title} {{Stimulated
  Raman adiabatic passage in a three-level superconducting circuit}},\ }\href
  {https://doi.org/10.1038/ncomms10628} {\bibfield  {journal} {\bibinfo
  {journal} {Nature Communications}\ }\textbf {\bibinfo {volume} {7}},\
  \bibinfo {pages} {10628} (\bibinfo {year} {2016})}\BibitemShut {NoStop}%
\bibitem [{\citenamefont {Falci}\ \emph {et~al.}(2017)\citenamefont {Falci},
  \citenamefont {{Di Stefano}}, \citenamefont {Ridolfo}, \citenamefont
  {D'Arrigo}, \citenamefont {Paraoanu},\ and\ \citenamefont
  {Paladino}}]{Falci2017}%
  \BibitemOpen
  \bibfield  {author} {\bibinfo {author} {\bibfnamefont {G.}~\bibnamefont
  {Falci}}, \bibinfo {author} {\bibfnamefont {P.~G.}\ \bibnamefont {{Di
  Stefano}}}, \bibinfo {author} {\bibfnamefont {A.}~\bibnamefont {Ridolfo}},
  \bibinfo {author} {\bibfnamefont {A.}~\bibnamefont {D'Arrigo}}, \bibinfo
  {author} {\bibfnamefont {G.~S.}\ \bibnamefont {Paraoanu}},\ and\ \bibinfo
  {author} {\bibfnamefont {E.}~\bibnamefont {Paladino}},\ }\bibfield  {title}
  {\bibinfo {title} {{Advances in quantum control of three-level
  superconducting circuit architectures}},\ }\href
  {https://doi.org/10.1002/prop.201600077} {\bibfield  {journal} {\bibinfo
  {journal} {Fortschritte der Physik}\ }\textbf {\bibinfo {volume} {65}},\
  \bibinfo {pages} {1600077} (\bibinfo {year} {2017})}\BibitemShut {NoStop}%
\bibitem [{\citenamefont {Dimer}\ \emph {et~al.}(2007)\citenamefont {Dimer},
  \citenamefont {Estienne}, \citenamefont {Parkins},\ and\ \citenamefont
  {Carmichael}}]{Dimer2007}%
  \BibitemOpen
  \bibfield  {author} {\bibinfo {author} {\bibfnamefont {F.}~\bibnamefont
  {Dimer}}, \bibinfo {author} {\bibfnamefont {B.}~\bibnamefont {Estienne}},
  \bibinfo {author} {\bibfnamefont {A.~S.}\ \bibnamefont {Parkins}},\ and\
  \bibinfo {author} {\bibfnamefont {H.~J.}\ \bibnamefont {Carmichael}},\
  }\bibfield  {title} {\bibinfo {title} {{Proposed realization of the
  Dicke-model quantum phase transition in an optical cavity QED system}},\
  }\href {https://doi.org/10.1103/PhysRevA.75.013804} {\bibfield  {journal}
  {\bibinfo  {journal} {Physical Review A}\ }\textbf {\bibinfo {volume} {75}},\
  \bibinfo {pages} {013804} (\bibinfo {year} {2007})}\BibitemShut {NoStop}%
\bibitem [{\citenamefont {Sun}\ \emph {et~al.}(2018)\citenamefont {Sun},
  \citenamefont {Zhang}, \citenamefont {Fischer}, \citenamefont {Burek},
  \citenamefont {Dory}, \citenamefont {Lagoudakis}, \citenamefont {Tzeng},
  \citenamefont {Radulaski}, \citenamefont {Kelaita}, \citenamefont
  {Safavi-Naeini}, \citenamefont {Shen}, \citenamefont {Melosh}, \citenamefont
  {Chu}, \citenamefont {Lon{\v{c}}ar},\ and\ \citenamefont
  {Vu{\v{c}}kovi{\'{c}}}}]{Sun2018}%
  \BibitemOpen
  \bibfield  {author} {\bibinfo {author} {\bibfnamefont {S.}~\bibnamefont
  {Sun}}, \bibinfo {author} {\bibfnamefont {J.~L.}\ \bibnamefont {Zhang}},
  \bibinfo {author} {\bibfnamefont {K.~A.}\ \bibnamefont {Fischer}}, \bibinfo
  {author} {\bibfnamefont {M.~J.}\ \bibnamefont {Burek}}, \bibinfo {author}
  {\bibfnamefont {C.}~\bibnamefont {Dory}}, \bibinfo {author} {\bibfnamefont
  {K.~G.}\ \bibnamefont {Lagoudakis}}, \bibinfo {author} {\bibfnamefont
  {Y.-K.}\ \bibnamefont {Tzeng}}, \bibinfo {author} {\bibfnamefont
  {M.}~\bibnamefont {Radulaski}}, \bibinfo {author} {\bibfnamefont
  {Y.}~\bibnamefont {Kelaita}}, \bibinfo {author} {\bibfnamefont
  {A.}~\bibnamefont {Safavi-Naeini}}, \bibinfo {author} {\bibfnamefont {Z.-X.}\
  \bibnamefont {Shen}}, \bibinfo {author} {\bibfnamefont {N.~A.}\ \bibnamefont
  {Melosh}}, \bibinfo {author} {\bibfnamefont {S.}~\bibnamefont {Chu}},
  \bibinfo {author} {\bibfnamefont {M.}~\bibnamefont {Lon{\v{c}}ar}},\ and\
  \bibinfo {author} {\bibfnamefont {J.}~\bibnamefont {Vu{\v{c}}kovi{\'{c}}}},\
  }\bibfield  {title} {\bibinfo {title} {{Cavity-Enhanced Raman Emission from a
  Single Color Center in a Solid}},\ }\href
  {https://doi.org/10.1103/PhysRevLett.121.083601} {\bibfield  {journal}
  {\bibinfo  {journal} {Physical Review Letters}\ }\textbf {\bibinfo {volume}
  {121}},\ \bibinfo {pages} {083601} (\bibinfo {year} {2018})}\BibitemShut
  {NoStop}%
\bibitem [{\citenamefont {Hennrich}\ \emph {et~al.}(2000)\citenamefont
  {Hennrich}, \citenamefont {Legero}, \citenamefont {Kuhn},\ and\ \citenamefont
  {Rempe}}]{Hennrich2000}%
  \BibitemOpen
  \bibfield  {author} {\bibinfo {author} {\bibfnamefont {M.}~\bibnamefont
  {Hennrich}}, \bibinfo {author} {\bibfnamefont {T.}~\bibnamefont {Legero}},
  \bibinfo {author} {\bibfnamefont {A.}~\bibnamefont {Kuhn}},\ and\ \bibinfo
  {author} {\bibfnamefont {G.}~\bibnamefont {Rempe}},\ }\bibfield  {title}
  {\bibinfo {title} {{Vacuum-Stimulated Raman Scattering Based on Adiabatic
  Passage in a High-Finesse Optical Cavity}},\ }\href
  {https://doi.org/10.1103/PhysRevLett.85.4872} {\bibfield  {journal} {\bibinfo
   {journal} {Physical Review Letters}\ }\textbf {\bibinfo {volume} {85}},\
  \bibinfo {pages} {4872} (\bibinfo {year} {2000})}\BibitemShut {NoStop}%
\bibitem [{\citenamefont {Kuhn}\ \emph {et~al.}(2002)\citenamefont {Kuhn},
  \citenamefont {Hennrich},\ and\ \citenamefont {Rempe}}]{Kuhn2002}%
  \BibitemOpen
  \bibfield  {author} {\bibinfo {author} {\bibfnamefont {A.}~\bibnamefont
  {Kuhn}}, \bibinfo {author} {\bibfnamefont {M.}~\bibnamefont {Hennrich}},\
  and\ \bibinfo {author} {\bibfnamefont {G.}~\bibnamefont {Rempe}},\ }\bibfield
   {title} {\bibinfo {title} {{Deterministic Single-Photon Source for
  Distributed Quantum Networking}},\ }\href
  {https://doi.org/10.1103/PhysRevLett.89.067901} {\bibfield  {journal}
  {\bibinfo  {journal} {Physical Review Letters}\ }\textbf {\bibinfo {volume}
  {89}},\ \bibinfo {pages} {067901} (\bibinfo {year} {2002})}\BibitemShut
  {NoStop}%
\bibitem [{\citenamefont {Sweeney}\ \emph {et~al.}(2014)\citenamefont
  {Sweeney}, \citenamefont {Carter}, \citenamefont {Bracker}, \citenamefont
  {Kim}, \citenamefont {Kim}, \citenamefont {Yang}, \citenamefont {Vora},
  \citenamefont {Brereton}, \citenamefont {Cleveland},\ and\ \citenamefont
  {Gammon}}]{Sweeney2014}%
  \BibitemOpen
  \bibfield  {author} {\bibinfo {author} {\bibfnamefont {T.~M.}\ \bibnamefont
  {Sweeney}}, \bibinfo {author} {\bibfnamefont {S.~G.}\ \bibnamefont {Carter}},
  \bibinfo {author} {\bibfnamefont {A.~S.}\ \bibnamefont {Bracker}}, \bibinfo
  {author} {\bibfnamefont {M.}~\bibnamefont {Kim}}, \bibinfo {author}
  {\bibfnamefont {C.~S.}\ \bibnamefont {Kim}}, \bibinfo {author} {\bibfnamefont
  {L.}~\bibnamefont {Yang}}, \bibinfo {author} {\bibfnamefont {P.~M.}\
  \bibnamefont {Vora}}, \bibinfo {author} {\bibfnamefont {P.~G.}\ \bibnamefont
  {Brereton}}, \bibinfo {author} {\bibfnamefont {E.~R.}\ \bibnamefont
  {Cleveland}},\ and\ \bibinfo {author} {\bibfnamefont {D.}~\bibnamefont
  {Gammon}},\ }\bibfield  {title} {\bibinfo {title} {{Cavity-stimulated Raman
  emission from a single quantum dot spin}},\ }\href
  {https://doi.org/10.1038/nphoton.2014.84} {\bibfield  {journal} {\bibinfo
  {journal} {Nature Photonics}\ }\textbf {\bibinfo {volume} {8}},\ \bibinfo
  {pages} {442} (\bibinfo {year} {2014})}\BibitemShut {NoStop}%
\bibitem [{\citenamefont {Orlando}\ \emph {et~al.}(2021)\citenamefont
  {Orlando}, \citenamefont {Franceschini}, \citenamefont {Muscas},
  \citenamefont {Pidkova}, \citenamefont {Bartoli}, \citenamefont {Rovere},\
  and\ \citenamefont {Tagliaferro}}]{Orlando2021}%
  \BibitemOpen
  \bibfield  {author} {\bibinfo {author} {\bibfnamefont {A.}~\bibnamefont
  {Orlando}}, \bibinfo {author} {\bibfnamefont {F.}~\bibnamefont
  {Franceschini}}, \bibinfo {author} {\bibfnamefont {C.}~\bibnamefont
  {Muscas}}, \bibinfo {author} {\bibfnamefont {S.}~\bibnamefont {Pidkova}},
  \bibinfo {author} {\bibfnamefont {M.}~\bibnamefont {Bartoli}}, \bibinfo
  {author} {\bibfnamefont {M.}~\bibnamefont {Rovere}},\ and\ \bibinfo {author}
  {\bibfnamefont {A.}~\bibnamefont {Tagliaferro}},\ }\bibfield  {title}
  {\bibinfo {title} {{A Comprehensive Review on Raman Spectroscopy
  Applications}},\ }\href {https://doi.org/10.3390/chemosensors9090262}
  {\bibfield  {journal} {\bibinfo  {journal} {Chemosensors}\ }\textbf {\bibinfo
  {volume} {9}},\ \bibinfo {pages} {262} (\bibinfo {year} {2021})}\BibitemShut
  {NoStop}%
\bibitem [{\citenamefont {Kudelski}(2008)}]{Kudelski2008}%
  \BibitemOpen
  \bibfield  {author} {\bibinfo {author} {\bibfnamefont {A.}~\bibnamefont
  {Kudelski}},\ }\bibfield  {title} {\bibinfo {title} {{Analytical applications
  of Raman spectroscopy}},\ }\href
  {https://doi.org/10.1016/j.talanta.2008.02.042} {\bibfield  {journal}
  {\bibinfo  {journal} {Talanta}\ }\textbf {\bibinfo {volume} {76}},\ \bibinfo
  {pages} {1} (\bibinfo {year} {2008})}\BibitemShut {NoStop}%
\bibitem [{\citenamefont {Pettinger}\ \emph {et~al.}(2012)\citenamefont
  {Pettinger}, \citenamefont {Schambach}, \citenamefont {Villag{\'{o}}mez},\
  and\ \citenamefont {Scott}}]{Pettinger2012}%
  \BibitemOpen
  \bibfield  {author} {\bibinfo {author} {\bibfnamefont {B.}~\bibnamefont
  {Pettinger}}, \bibinfo {author} {\bibfnamefont {P.}~\bibnamefont
  {Schambach}}, \bibinfo {author} {\bibfnamefont {C.~J.}\ \bibnamefont
  {Villag{\'{o}}mez}},\ and\ \bibinfo {author} {\bibfnamefont {N.}~\bibnamefont
  {Scott}},\ }\bibfield  {title} {\bibinfo {title} {{Tip-Enhanced Raman
  Spectroscopy: Near-Fields Acting on a Few Molecules}},\ }\href
  {https://doi.org/10.1146/annurev-physchem-032511-143807} {\bibfield
  {journal} {\bibinfo  {journal} {Annual Review of Physical Chemistry}\
  }\textbf {\bibinfo {volume} {63}},\ \bibinfo {pages} {379} (\bibinfo {year}
  {2012})}\BibitemShut {NoStop}%
\bibitem [{\citenamefont {Qian}\ and\ \citenamefont {Nie}(2008)}]{Qian2008}%
  \BibitemOpen
  \bibfield  {author} {\bibinfo {author} {\bibfnamefont {X.-M.}\ \bibnamefont
  {Qian}}\ and\ \bibinfo {author} {\bibfnamefont {S.~M.}\ \bibnamefont {Nie}},\
  }\bibfield  {title} {\bibinfo {title} {{Single-molecule and
  single-nanoparticle SERS: from fundamental mechanisms to biomedical
  applications}},\ }\href {https://doi.org/10.1039/b708839f} {\bibfield
  {journal} {\bibinfo  {journal} {Chemical Society Reviews}\ }\textbf {\bibinfo
  {volume} {37}},\ \bibinfo {pages} {912} (\bibinfo {year} {2008})}\BibitemShut
  {NoStop}%
\bibitem [{\citenamefont {Yang}\ and\ \citenamefont {Ying}(2011)}]{Yang2011}%
  \BibitemOpen
  \bibfield  {author} {\bibinfo {author} {\bibfnamefont {D.}~\bibnamefont
  {Yang}}\ and\ \bibinfo {author} {\bibfnamefont {Y.}~\bibnamefont {Ying}},\
  }\bibfield  {title} {\bibinfo {title} {{Applications of Raman Spectroscopy in
  Agricultural Products and Food Analysis: A Review}},\ }\href
  {https://doi.org/10.1080/05704928.2011.593216} {\bibfield  {journal}
  {\bibinfo  {journal} {Applied Spectroscopy Reviews}\ }\textbf {\bibinfo
  {volume} {46}},\ \bibinfo {pages} {539} (\bibinfo {year} {2011})}\BibitemShut
  {NoStop}%
\bibitem [{\citenamefont {{Poornima Parvathi}}\ \emph
  {et~al.}(2019)\citenamefont {{Poornima Parvathi}}, \citenamefont
  {Parimaladevi}, \citenamefont {Sathe},\ and\ \citenamefont
  {Mahalingam}}]{PoornimaParvathi2019}%
  \BibitemOpen
  \bibfield  {author} {\bibinfo {author} {\bibfnamefont {V.}~\bibnamefont
  {{Poornima Parvathi}}}, \bibinfo {author} {\bibfnamefont {R.}~\bibnamefont
  {Parimaladevi}}, \bibinfo {author} {\bibfnamefont {V.}~\bibnamefont
  {Sathe}},\ and\ \bibinfo {author} {\bibfnamefont {U.}~\bibnamefont
  {Mahalingam}},\ }\bibfield  {title} {\bibinfo {title} {{Graphene boosted
  silver nanoparticles as surface enhanced Raman spectroscopic sensors and
  photocatalysts for removal of standard and industrial dye contaminants}},\
  }\href {https://doi.org/10.1016/j.snb.2018.11.007} {\bibfield  {journal}
  {\bibinfo  {journal} {Sensors and Actuators B: Chemical}\ }\textbf {\bibinfo
  {volume} {281}},\ \bibinfo {pages} {679} (\bibinfo {year}
  {2019})}\BibitemShut {NoStop}%
\bibitem [{\citenamefont {Langer}\ \emph {et~al.}(2020)\citenamefont {Langer},
  \citenamefont {{Jimenez de Aberasturi}}, \citenamefont {Aizpurua},
  \citenamefont {Alvarez-Puebla}, \citenamefont {Augui{\'{e}}}, \citenamefont
  {Baumberg}, \citenamefont {Bazan}, \citenamefont {Bell}, \citenamefont
  {Boisen}, \citenamefont {Brolo}, \citenamefont {Choo}, \citenamefont
  {Cialla-May}, \citenamefont {Deckert}, \citenamefont {Fabris}, \citenamefont
  {Faulds}, \citenamefont {{Garc{\'{i}}a de Abajo}}, \citenamefont {Goodacre},
  \citenamefont {Graham}, \citenamefont {Haes}, \citenamefont {Haynes},
  \citenamefont {Huck}, \citenamefont {Itoh}, \citenamefont {K{\"{a}}ll},
  \citenamefont {Kneipp}, \citenamefont {Kotov}, \citenamefont {Kuang},
  \citenamefont {{Le Ru}}, \citenamefont {Lee}, \citenamefont {Li},
  \citenamefont {Ling}, \citenamefont {Maier}, \citenamefont
  {Mayerh{\"{o}}fer}, \citenamefont {Moskovits}, \citenamefont {Murakoshi},
  \citenamefont {Nam}, \citenamefont {Nie}, \citenamefont {Ozaki},
  \citenamefont {Pastoriza-Santos}, \citenamefont {Perez-Juste}, \citenamefont
  {Popp}, \citenamefont {Pucci}, \citenamefont {Reich}, \citenamefont {Ren},
  \citenamefont {Schatz}, \citenamefont {Shegai}, \citenamefont
  {Schl{\"{u}}cker}, \citenamefont {Tay}, \citenamefont {Thomas}, \citenamefont
  {Tian}, \citenamefont {{Van Duyne}}, \citenamefont {Vo-Dinh}, \citenamefont
  {Wang}, \citenamefont {Willets}, \citenamefont {Xu}, \citenamefont {Xu},
  \citenamefont {Xu}, \citenamefont {Yamamoto}, \citenamefont {Zhao},\ and\
  \citenamefont {Liz-Marz{\'{a}}n}}]{Langer2020}%
  \BibitemOpen
  \bibfield  {author} {\bibinfo {author} {\bibfnamefont {J.}~\bibnamefont
  {Langer}}, \bibinfo {author} {\bibfnamefont {D.}~\bibnamefont {{Jimenez de
  Aberasturi}}}, \bibinfo {author} {\bibfnamefont {J.}~\bibnamefont
  {Aizpurua}}, \bibinfo {author} {\bibfnamefont {R.~A.}\ \bibnamefont
  {Alvarez-Puebla}}, \bibinfo {author} {\bibfnamefont {B.}~\bibnamefont
  {Augui{\'{e}}}}, \bibinfo {author} {\bibfnamefont {J.~J.}\ \bibnamefont
  {Baumberg}}, \bibinfo {author} {\bibfnamefont {G.~C.}\ \bibnamefont {Bazan}},
  \bibinfo {author} {\bibfnamefont {S.~E.~J.}\ \bibnamefont {Bell}}, \bibinfo
  {author} {\bibfnamefont {A.}~\bibnamefont {Boisen}}, \bibinfo {author}
  {\bibfnamefont {A.~G.}\ \bibnamefont {Brolo}}, \bibinfo {author}
  {\bibfnamefont {J.}~\bibnamefont {Choo}}, \bibinfo {author} {\bibfnamefont
  {D.}~\bibnamefont {Cialla-May}}, \bibinfo {author} {\bibfnamefont
  {V.}~\bibnamefont {Deckert}}, \bibinfo {author} {\bibfnamefont
  {L.}~\bibnamefont {Fabris}}, \bibinfo {author} {\bibfnamefont
  {K.}~\bibnamefont {Faulds}}, \bibinfo {author} {\bibfnamefont {F.~J.}\
  \bibnamefont {{Garc{\'{i}}a de Abajo}}}, \bibinfo {author} {\bibfnamefont
  {R.}~\bibnamefont {Goodacre}}, \bibinfo {author} {\bibfnamefont
  {D.}~\bibnamefont {Graham}}, \bibinfo {author} {\bibfnamefont {A.~J.}\
  \bibnamefont {Haes}}, \bibinfo {author} {\bibfnamefont {C.~L.}\ \bibnamefont
  {Haynes}}, \bibinfo {author} {\bibfnamefont {C.}~\bibnamefont {Huck}},
  \bibinfo {author} {\bibfnamefont {T.}~\bibnamefont {Itoh}}, \bibinfo {author}
  {\bibfnamefont {M.}~\bibnamefont {K{\"{a}}ll}}, \bibinfo {author}
  {\bibfnamefont {J.}~\bibnamefont {Kneipp}}, \bibinfo {author} {\bibfnamefont
  {N.~A.}\ \bibnamefont {Kotov}}, \bibinfo {author} {\bibfnamefont
  {H.}~\bibnamefont {Kuang}}, \bibinfo {author} {\bibfnamefont {E.~C.}\
  \bibnamefont {{Le Ru}}}, \bibinfo {author} {\bibfnamefont {H.~K.}\
  \bibnamefont {Lee}}, \bibinfo {author} {\bibfnamefont {J.-F.}\ \bibnamefont
  {Li}}, \bibinfo {author} {\bibfnamefont {X.~Y.}\ \bibnamefont {Ling}},
  \bibinfo {author} {\bibfnamefont {S.~A.}\ \bibnamefont {Maier}}, \bibinfo
  {author} {\bibfnamefont {T.}~\bibnamefont {Mayerh{\"{o}}fer}}, \bibinfo
  {author} {\bibfnamefont {M.}~\bibnamefont {Moskovits}}, \bibinfo {author}
  {\bibfnamefont {K.}~\bibnamefont {Murakoshi}}, \bibinfo {author}
  {\bibfnamefont {J.-M.}\ \bibnamefont {Nam}}, \bibinfo {author} {\bibfnamefont
  {S.}~\bibnamefont {Nie}}, \bibinfo {author} {\bibfnamefont {Y.}~\bibnamefont
  {Ozaki}}, \bibinfo {author} {\bibfnamefont {I.}~\bibnamefont
  {Pastoriza-Santos}}, \bibinfo {author} {\bibfnamefont {J.}~\bibnamefont
  {Perez-Juste}}, \bibinfo {author} {\bibfnamefont {J.}~\bibnamefont {Popp}},
  \bibinfo {author} {\bibfnamefont {A.}~\bibnamefont {Pucci}}, \bibinfo
  {author} {\bibfnamefont {S.}~\bibnamefont {Reich}}, \bibinfo {author}
  {\bibfnamefont {B.}~\bibnamefont {Ren}}, \bibinfo {author} {\bibfnamefont
  {G.~C.}\ \bibnamefont {Schatz}}, \bibinfo {author} {\bibfnamefont
  {T.}~\bibnamefont {Shegai}}, \bibinfo {author} {\bibfnamefont
  {S.}~\bibnamefont {Schl{\"{u}}cker}}, \bibinfo {author} {\bibfnamefont
  {L.-L.}\ \bibnamefont {Tay}}, \bibinfo {author} {\bibfnamefont {K.~G.}\
  \bibnamefont {Thomas}}, \bibinfo {author} {\bibfnamefont {Z.-Q.}\
  \bibnamefont {Tian}}, \bibinfo {author} {\bibfnamefont {R.~P.}\ \bibnamefont
  {{Van Duyne}}}, \bibinfo {author} {\bibfnamefont {T.}~\bibnamefont
  {Vo-Dinh}}, \bibinfo {author} {\bibfnamefont {Y.}~\bibnamefont {Wang}},
  \bibinfo {author} {\bibfnamefont {K.~A.}\ \bibnamefont {Willets}}, \bibinfo
  {author} {\bibfnamefont {C.}~\bibnamefont {Xu}}, \bibinfo {author}
  {\bibfnamefont {H.}~\bibnamefont {Xu}}, \bibinfo {author} {\bibfnamefont
  {Y.}~\bibnamefont {Xu}}, \bibinfo {author} {\bibfnamefont {Y.~S.}\
  \bibnamefont {Yamamoto}}, \bibinfo {author} {\bibfnamefont {B.}~\bibnamefont
  {Zhao}},\ and\ \bibinfo {author} {\bibfnamefont {L.~M.}\ \bibnamefont
  {Liz-Marz{\'{a}}n}},\ }\bibfield  {title} {\bibinfo {title} {{Present and
  Future of Surface-Enhanced Raman Scattering}},\ }\href
  {https://doi.org/10.1021/acsnano.9b04224} {\bibfield  {journal} {\bibinfo
  {journal} {ACS Nano}\ }\textbf {\bibinfo {volume} {14}},\ \bibinfo {pages}
  {28} (\bibinfo {year} {2020})}\BibitemShut {NoStop}%
\bibitem [{\citenamefont {Kneipp}\ \emph {et~al.}(1997)\citenamefont {Kneipp},
  \citenamefont {Wang}, \citenamefont {Kneipp}, \citenamefont {Perelman},
  \citenamefont {Itzkan}, \citenamefont {Dasari},\ and\ \citenamefont
  {Feld}}]{Kneipp1997}%
  \BibitemOpen
  \bibfield  {author} {\bibinfo {author} {\bibfnamefont {K.}~\bibnamefont
  {Kneipp}}, \bibinfo {author} {\bibfnamefont {Y.}~\bibnamefont {Wang}},
  \bibinfo {author} {\bibfnamefont {H.}~\bibnamefont {Kneipp}}, \bibinfo
  {author} {\bibfnamefont {L.~T.}\ \bibnamefont {Perelman}}, \bibinfo {author}
  {\bibfnamefont {I.}~\bibnamefont {Itzkan}}, \bibinfo {author} {\bibfnamefont
  {R.~R.}\ \bibnamefont {Dasari}},\ and\ \bibinfo {author} {\bibfnamefont
  {M.~S.}\ \bibnamefont {Feld}},\ }\bibfield  {title} {\bibinfo {title}
  {{Single Molecule Detection Using Surface-Enhanced Raman Scattering
  (SERS)}},\ }\href {https://doi.org/10.1103/PhysRevLett.78.1667} {\bibfield
  {journal} {\bibinfo  {journal} {Physical Review Letters}\ }\textbf {\bibinfo
  {volume} {78}},\ \bibinfo {pages} {1667} (\bibinfo {year}
  {1997})}\BibitemShut {NoStop}%
\bibitem [{\citenamefont {Nie}\ and\ \citenamefont {Emory}(1997)}]{Nie1997}%
  \BibitemOpen
  \bibfield  {author} {\bibinfo {author} {\bibfnamefont {S.}~\bibnamefont
  {Nie}}\ and\ \bibinfo {author} {\bibfnamefont {S.~R.}\ \bibnamefont
  {Emory}},\ }\bibfield  {title} {\bibinfo {title} {{Probing Single Molecules
  and Single Nanoparticles by Surface-Enhanced Raman Scattering}},\ }\href
  {https://doi.org/10.1126/science.275.5303.1102} {\bibfield  {journal}
  {\bibinfo  {journal} {Science}\ }\textbf {\bibinfo {volume} {275}},\ \bibinfo
  {pages} {1102} (\bibinfo {year} {1997})}\BibitemShut {NoStop}%
\bibitem [{\citenamefont {Zhang}\ \emph {et~al.}(2013)\citenamefont {Zhang},
  \citenamefont {Zhang}, \citenamefont {Dong}, \citenamefont {Jiang},
  \citenamefont {Zhang}, \citenamefont {Chen}, \citenamefont {Zhang},
  \citenamefont {Liao}, \citenamefont {Aizpurua}, \citenamefont {Luo},
  \citenamefont {Yang},\ and\ \citenamefont {Hou}}]{Zhang2013}%
  \BibitemOpen
  \bibfield  {author} {\bibinfo {author} {\bibfnamefont {R.}~\bibnamefont
  {Zhang}}, \bibinfo {author} {\bibfnamefont {Y.}~\bibnamefont {Zhang}},
  \bibinfo {author} {\bibfnamefont {Z.~C.}\ \bibnamefont {Dong}}, \bibinfo
  {author} {\bibfnamefont {S.}~\bibnamefont {Jiang}}, \bibinfo {author}
  {\bibfnamefont {C.}~\bibnamefont {Zhang}}, \bibinfo {author} {\bibfnamefont
  {L.~G.}\ \bibnamefont {Chen}}, \bibinfo {author} {\bibfnamefont
  {L.}~\bibnamefont {Zhang}}, \bibinfo {author} {\bibfnamefont
  {Y.}~\bibnamefont {Liao}}, \bibinfo {author} {\bibfnamefont {J.}~\bibnamefont
  {Aizpurua}}, \bibinfo {author} {\bibfnamefont {Y.}~\bibnamefont {Luo}},
  \bibinfo {author} {\bibfnamefont {J.~L.}\ \bibnamefont {Yang}},\ and\
  \bibinfo {author} {\bibfnamefont {J.~G.}\ \bibnamefont {Hou}},\ }\bibfield
  {title} {\bibinfo {title} {{Chemical mapping of a single molecule by
  plasmon-enhanced Raman scattering}},\ }\href
  {https://doi.org/10.1038/nature12151} {\bibfield  {journal} {\bibinfo
  {journal} {Nature}\ }\textbf {\bibinfo {volume} {498}},\ \bibinfo {pages}
  {82} (\bibinfo {year} {2013})}\BibitemShut {NoStop}%
\bibitem [{\citenamefont {Zhu}\ and\ \citenamefont {Crozier}(2014)}]{Zhu2014}%
  \BibitemOpen
  \bibfield  {author} {\bibinfo {author} {\bibfnamefont {W.}~\bibnamefont
  {Zhu}}\ and\ \bibinfo {author} {\bibfnamefont {K.~B.}\ \bibnamefont
  {Crozier}},\ }\bibfield  {title} {\bibinfo {title} {{Quantum mechanical limit
  to plasmonic enhancement as observed by surface-enhanced Raman scattering}},\
  }\href {https://doi.org/10.1038/ncomms6228} {\bibfield  {journal} {\bibinfo
  {journal} {Nature Communications}\ }\textbf {\bibinfo {volume} {5}},\
  \bibinfo {pages} {5228} (\bibinfo {year} {2014})}\BibitemShut {NoStop}%
\bibitem [{\citenamefont {Shalabney}\ \emph {et~al.}(2015)\citenamefont
  {Shalabney}, \citenamefont {George}, \citenamefont {Hutchison}, \citenamefont
  {Pupillo}, \citenamefont {Genet},\ and\ \citenamefont
  {Ebbesen}}]{Shalabney2015}%
  \BibitemOpen
  \bibfield  {author} {\bibinfo {author} {\bibfnamefont {A.}~\bibnamefont
  {Shalabney}}, \bibinfo {author} {\bibfnamefont {J.}~\bibnamefont {George}},
  \bibinfo {author} {\bibfnamefont {J.}~\bibnamefont {Hutchison}}, \bibinfo
  {author} {\bibfnamefont {G.}~\bibnamefont {Pupillo}}, \bibinfo {author}
  {\bibfnamefont {C.}~\bibnamefont {Genet}},\ and\ \bibinfo {author}
  {\bibfnamefont {T.~W.}\ \bibnamefont {Ebbesen}},\ }\bibfield  {title}
  {\bibinfo {title} {{Coherent coupling of molecular resonators with a
  microcavity mode}},\ }\href {https://doi.org/10.1038/ncomms6981} {\bibfield
  {journal} {\bibinfo  {journal} {Nature Communications}\ }\textbf {\bibinfo
  {volume} {6}},\ \bibinfo {pages} {5981} (\bibinfo {year} {2015})}\BibitemShut
  {NoStop}%
\bibitem [{\citenamefont {Schmidt}\ \emph {et~al.}(2016)\citenamefont
  {Schmidt}, \citenamefont {Esteban}, \citenamefont {Gonz{\'{a}}lez-Tudela},
  \citenamefont {Giedke},\ and\ \citenamefont {Aizpurua}}]{Schmidt2016}%
  \BibitemOpen
  \bibfield  {author} {\bibinfo {author} {\bibfnamefont {M.~K.}\ \bibnamefont
  {Schmidt}}, \bibinfo {author} {\bibfnamefont {R.}~\bibnamefont {Esteban}},
  \bibinfo {author} {\bibfnamefont {A.}~\bibnamefont {Gonz{\'{a}}lez-Tudela}},
  \bibinfo {author} {\bibfnamefont {G.}~\bibnamefont {Giedke}},\ and\ \bibinfo
  {author} {\bibfnamefont {J.}~\bibnamefont {Aizpurua}},\ }\bibfield  {title}
  {\bibinfo {title} {{Quantum Mechanical Description of Raman Scattering from
  Molecules in Plasmonic Cavities}},\ }\href
  {https://doi.org/10.1021/acsnano.6b02484} {\bibfield  {journal} {\bibinfo
  {journal} {ACS Nano}\ }\textbf {\bibinfo {volume} {10}},\ \bibinfo {pages}
  {6291} (\bibinfo {year} {2016})}\BibitemShut {NoStop}%
\bibitem [{\citenamefont {Roelli}\ \emph {et~al.}(2016)\citenamefont {Roelli},
  \citenamefont {Galland}, \citenamefont {Piro},\ and\ \citenamefont
  {Kippenberg}}]{Roelli2016}%
  \BibitemOpen
  \bibfield  {author} {\bibinfo {author} {\bibfnamefont {P.}~\bibnamefont
  {Roelli}}, \bibinfo {author} {\bibfnamefont {C.}~\bibnamefont {Galland}},
  \bibinfo {author} {\bibfnamefont {N.}~\bibnamefont {Piro}},\ and\ \bibinfo
  {author} {\bibfnamefont {T.~J.}\ \bibnamefont {Kippenberg}},\ }\bibfield
  {title} {\bibinfo {title} {{Molecular cavity optomechanics as a theory of
  plasmon-enhanced Raman scattering}},\ }\href
  {https://doi.org/10.1038/nnano.2015.264} {\bibfield  {journal} {\bibinfo
  {journal} {Nature Nanotechnology}\ }\textbf {\bibinfo {volume} {11}},\
  \bibinfo {pages} {164} (\bibinfo {year} {2016})}\BibitemShut {NoStop}%
\bibitem [{\citenamefont {{Kamandar Dezfouli}}\ and\ \citenamefont
  {Hughes}(2017)}]{KamandarDezfouli2017}%
  \BibitemOpen
  \bibfield  {author} {\bibinfo {author} {\bibfnamefont {M.}~\bibnamefont
  {{Kamandar Dezfouli}}}\ and\ \bibinfo {author} {\bibfnamefont
  {S.}~\bibnamefont {Hughes}},\ }\bibfield  {title} {\bibinfo {title} {{Quantum
  Optics Model of Surface-Enhanced Raman Spectroscopy for Arbitrarily Shaped
  Plasmonic Resonators}},\ }\href
  {https://doi.org/10.1021/acsphotonics.7b00157} {\bibfield  {journal}
  {\bibinfo  {journal} {ACS Photonics}\ }\textbf {\bibinfo {volume} {4}},\
  \bibinfo {pages} {1045} (\bibinfo {year} {2017})}\BibitemShut {NoStop}%
\bibitem [{\citenamefont {Dezfouli}\ \emph {et~al.}(2019)\citenamefont
  {Dezfouli}, \citenamefont {Gordon},\ and\ \citenamefont
  {Hughes}}]{Dezfouli2019}%
  \BibitemOpen
  \bibfield  {author} {\bibinfo {author} {\bibfnamefont {M.~K.}\ \bibnamefont
  {Dezfouli}}, \bibinfo {author} {\bibfnamefont {R.}~\bibnamefont {Gordon}},\
  and\ \bibinfo {author} {\bibfnamefont {S.}~\bibnamefont {Hughes}},\
  }\bibfield  {title} {\bibinfo {title} {{Molecular Optomechanics in the
  Anharmonic Cavity-QED Regime Using Hybrid Metal–Dielectric Cavity Modes}},\
  }\href {https://doi.org/10.1021/acsphotonics.8b01091} {\bibfield  {journal}
  {\bibinfo  {journal} {ACS Photonics}\ }\textbf {\bibinfo {volume} {6}},\
  \bibinfo {pages} {1400} (\bibinfo {year} {2019})}\BibitemShut {NoStop}%
\bibitem [{\citenamefont {Hughes}\ \emph {et~al.}(2021)\citenamefont {Hughes},
  \citenamefont {Settineri}, \citenamefont {Savasta},\ and\ \citenamefont
  {Nori}}]{Hughes2021}%
  \BibitemOpen
  \bibfield  {author} {\bibinfo {author} {\bibfnamefont {S.}~\bibnamefont
  {Hughes}}, \bibinfo {author} {\bibfnamefont {A.}~\bibnamefont {Settineri}},
  \bibinfo {author} {\bibfnamefont {S.}~\bibnamefont {Savasta}},\ and\ \bibinfo
  {author} {\bibfnamefont {F.}~\bibnamefont {Nori}},\ }\bibfield  {title}
  {\bibinfo {title} {{Resonant Raman scattering of single molecules under
  simultaneous strong cavity coupling and ultrastrong optomechanical coupling
  in plasmonic resonators: Phonon-dressed polaritons}},\ }\href
  {https://doi.org/10.1103/PhysRevB.104.045431} {\bibfield  {journal} {\bibinfo
   {journal} {Physical Review B}\ }\textbf {\bibinfo {volume} {104}},\ \bibinfo
  {pages} {045431} (\bibinfo {year} {2021})}\BibitemShut {NoStop}%
\bibitem [{\citenamefont {Roelli}\ \emph {et~al.}(2020)\citenamefont {Roelli},
  \citenamefont {Martin-Cano}, \citenamefont {Kippenberg},\ and\ \citenamefont
  {Galland}}]{Roelli2020}%
  \BibitemOpen
  \bibfield  {author} {\bibinfo {author} {\bibfnamefont {P.}~\bibnamefont
  {Roelli}}, \bibinfo {author} {\bibfnamefont {D.}~\bibnamefont {Martin-Cano}},
  \bibinfo {author} {\bibfnamefont {T.~J.}\ \bibnamefont {Kippenberg}},\ and\
  \bibinfo {author} {\bibfnamefont {C.}~\bibnamefont {Galland}},\ }\bibfield
  {title} {\bibinfo {title} {{Molecular Platform for Frequency Upconversion at
  the Single-Photon Level}},\ }\href
  {https://doi.org/10.1103/PhysRevX.10.031057} {\bibfield  {journal} {\bibinfo
  {journal} {Physical Review X}\ }\textbf {\bibinfo {volume} {10}},\ \bibinfo
  {pages} {031057} (\bibinfo {year} {2020})}\BibitemShut {NoStop}%
\bibitem [{\citenamefont {Gurlek}\ \emph {et~al.}(2021)\citenamefont {Gurlek},
  \citenamefont {Sandoghdar},\ and\ \citenamefont {Martin-Cano}}]{Gurlek2021}%
  \BibitemOpen
  \bibfield  {author} {\bibinfo {author} {\bibfnamefont {B.}~\bibnamefont
  {Gurlek}}, \bibinfo {author} {\bibfnamefont {V.}~\bibnamefont {Sandoghdar}},\
  and\ \bibinfo {author} {\bibfnamefont {D.}~\bibnamefont {Martin-Cano}},\
  }\bibfield  {title} {\bibinfo {title} {{Engineering Long-Lived Vibrational
  States for an Organic Molecule}},\ }\href
  {https://doi.org/10.1103/PhysRevLett.127.123603} {\bibfield  {journal}
  {\bibinfo  {journal} {Physical Review Letters}\ }\textbf {\bibinfo {volume}
  {127}},\ \bibinfo {pages} {123603} (\bibinfo {year} {2021})}\BibitemShut
  {NoStop}%
\bibitem [{\citenamefont {Scully}\ and\ \citenamefont
  {Zubairy}(2002)}]{scully_book02a}%
  \BibitemOpen
  \bibfield  {author} {\bibinfo {author} {\bibfnamefont {M.~O.}\ \bibnamefont
  {Scully}}\ and\ \bibinfo {author} {\bibfnamefont {M.~S.}\ \bibnamefont
  {Zubairy}},\ }\href@noop {} {\emph {\bibinfo {title} {Quantum optics}}}\
  (\bibinfo  {publisher} {Cambridge University Press},\ \bibinfo {year}
  {2002})\BibitemShut {NoStop}%
\bibitem [{\citenamefont {Agarwal}(2012)}]{agarwal_book12a}%
  \BibitemOpen
  \bibfield  {author} {\bibinfo {author} {\bibfnamefont {G.~S.}\ \bibnamefont
  {Agarwal}},\ }\href@noop {} {\emph {\bibinfo {title} {Quantum optics}}}\
  (\bibinfo  {publisher} {Cambridge University Press},\ \bibinfo {year}
  {2012})\BibitemShut {NoStop}%
\bibitem [{\citenamefont {{Frisk Kockum}}\ \emph {et~al.}(2019)\citenamefont
  {{Frisk Kockum}}, \citenamefont {Miranowicz}, \citenamefont {{De Liberato}},
  \citenamefont {Savasta},\ and\ \citenamefont {Nori}}]{FriskKockum2019}%
  \BibitemOpen
  \bibfield  {author} {\bibinfo {author} {\bibfnamefont {A.}~\bibnamefont
  {{Frisk Kockum}}}, \bibinfo {author} {\bibfnamefont {A.}~\bibnamefont
  {Miranowicz}}, \bibinfo {author} {\bibfnamefont {S.}~\bibnamefont {{De
  Liberato}}}, \bibinfo {author} {\bibfnamefont {S.}~\bibnamefont {Savasta}},\
  and\ \bibinfo {author} {\bibfnamefont {F.}~\bibnamefont {Nori}},\ }\bibfield
  {title} {\bibinfo {title} {{Ultrastrong coupling between light and matter}},\
  }\href {https://doi.org/10.1038/s42254-018-0006-2} {\bibfield  {journal}
  {\bibinfo  {journal} {Nature Reviews Physics}\ }\textbf {\bibinfo {volume}
  {1}},\ \bibinfo {pages} {19} (\bibinfo {year} {2019})}\BibitemShut {NoStop}%
\bibitem [{\citenamefont {Forn-D{\'{i}}az}\ \emph {et~al.}(2019)\citenamefont
  {Forn-D{\'{i}}az}, \citenamefont {Lamata}, \citenamefont {Rico},
  \citenamefont {Kono},\ and\ \citenamefont {Solano}}]{Forn-Diaz2019}%
  \BibitemOpen
  \bibfield  {author} {\bibinfo {author} {\bibfnamefont {P.}~\bibnamefont
  {Forn-D{\'{i}}az}}, \bibinfo {author} {\bibfnamefont {L.}~\bibnamefont
  {Lamata}}, \bibinfo {author} {\bibfnamefont {E.}~\bibnamefont {Rico}},
  \bibinfo {author} {\bibfnamefont {J.}~\bibnamefont {Kono}},\ and\ \bibinfo
  {author} {\bibfnamefont {E.}~\bibnamefont {Solano}},\ }\bibfield  {title}
  {\bibinfo {title} {{Ultrastrong coupling regimes of light-matter
  interaction}},\ }\href {https://doi.org/10.1103/RevModPhys.91.025005}
  {\bibfield  {journal} {\bibinfo  {journal} {Reviews of Modern Physics}\
  }\textbf {\bibinfo {volume} {91}},\ \bibinfo {pages} {025005} (\bibinfo
  {year} {2019})}\BibitemShut {NoStop}%
\bibitem [{\citenamefont {Niemczyk}\ \emph {et~al.}(2010)\citenamefont
  {Niemczyk}, \citenamefont {Deppe}, \citenamefont {Huebl}, \citenamefont
  {Menzel}, \citenamefont {Hocke}, \citenamefont {Schwarz}, \citenamefont
  {Garcia-Ripoll}, \citenamefont {Zueco}, \citenamefont {H{\"{u}}mmer},
  \citenamefont {Solano}, \citenamefont {Marx},\ and\ \citenamefont
  {Gross}}]{Niemczyk2010}%
  \BibitemOpen
  \bibfield  {author} {\bibinfo {author} {\bibfnamefont {T.}~\bibnamefont
  {Niemczyk}}, \bibinfo {author} {\bibfnamefont {F.}~\bibnamefont {Deppe}},
  \bibinfo {author} {\bibfnamefont {H.}~\bibnamefont {Huebl}}, \bibinfo
  {author} {\bibfnamefont {E.~P.}\ \bibnamefont {Menzel}}, \bibinfo {author}
  {\bibfnamefont {F.}~\bibnamefont {Hocke}}, \bibinfo {author} {\bibfnamefont
  {M.~J.}\ \bibnamefont {Schwarz}}, \bibinfo {author} {\bibfnamefont {J.~J.}\
  \bibnamefont {Garcia-Ripoll}}, \bibinfo {author} {\bibfnamefont
  {D.}~\bibnamefont {Zueco}}, \bibinfo {author} {\bibfnamefont
  {T.}~\bibnamefont {H{\"{u}}mmer}}, \bibinfo {author} {\bibfnamefont
  {E.}~\bibnamefont {Solano}}, \bibinfo {author} {\bibfnamefont
  {A.}~\bibnamefont {Marx}},\ and\ \bibinfo {author} {\bibfnamefont
  {R.}~\bibnamefont {Gross}},\ }\bibfield  {title} {\bibinfo {title} {{Circuit
  quantum electrodynamics in the ultrastrong-coupling regime}},\ }\href
  {https://doi.org/10.1038/nphys1730} {\bibfield  {journal} {\bibinfo
  {journal} {Nature Physics}\ }\textbf {\bibinfo {volume} {6}},\ \bibinfo
  {pages} {772} (\bibinfo {year} {2010})}\BibitemShut {NoStop}%
\bibitem [{\citenamefont {Ma}\ and\ \citenamefont {Law}(2015)}]{Ma2015}%
  \BibitemOpen
  \bibfield  {author} {\bibinfo {author} {\bibfnamefont {K.~K.~W.}\
  \bibnamefont {Ma}}\ and\ \bibinfo {author} {\bibfnamefont {C.~K.}\
  \bibnamefont {Law}},\ }\bibfield  {title} {\bibinfo {title} {{Three-photon
  resonance and adiabatic passage in the large-detuning Rabi model}},\ }\href
  {https://doi.org/10.1103/PhysRevA.92.023842} {\bibfield  {journal} {\bibinfo
  {journal} {Physical Review A}\ }\textbf {\bibinfo {volume} {92}},\ \bibinfo
  {pages} {023842} (\bibinfo {year} {2015})}\BibitemShut {NoStop}%
\bibitem [{\citenamefont {Garziano}\ \emph {et~al.}(2015)\citenamefont
  {Garziano}, \citenamefont {Stassi}, \citenamefont {Macr{\`{i}}},
  \citenamefont {Kockum}, \citenamefont {Savasta},\ and\ \citenamefont
  {Nori}}]{Garziano2015}%
  \BibitemOpen
  \bibfield  {author} {\bibinfo {author} {\bibfnamefont {L.}~\bibnamefont
  {Garziano}}, \bibinfo {author} {\bibfnamefont {R.}~\bibnamefont {Stassi}},
  \bibinfo {author} {\bibfnamefont {V.}~\bibnamefont {Macr{\`{i}}}}, \bibinfo
  {author} {\bibfnamefont {A.~F.}\ \bibnamefont {Kockum}}, \bibinfo {author}
  {\bibfnamefont {S.}~\bibnamefont {Savasta}},\ and\ \bibinfo {author}
  {\bibfnamefont {F.}~\bibnamefont {Nori}},\ }\bibfield  {title} {\bibinfo
  {title} {{Multiphoton quantum Rabi oscillations in ultrastrong cavity QED}},\
  }\href {https://doi.org/10.1103/PhysRevA.92.063830} {\bibfield  {journal}
  {\bibinfo  {journal} {Physical Review A}\ }\textbf {\bibinfo {volume} {92}},\
  \bibinfo {pages} {063830} (\bibinfo {year} {2015})}\BibitemShut {NoStop}%
\bibitem [{\citenamefont {Garziano}\ \emph {et~al.}(2016)\citenamefont
  {Garziano}, \citenamefont {Macr{\`{i}}}, \citenamefont {Stassi},
  \citenamefont {{Di Stefano}}, \citenamefont {Nori},\ and\ \citenamefont
  {Savasta}}]{Garziano2016}%
  \BibitemOpen
  \bibfield  {author} {\bibinfo {author} {\bibfnamefont {L.}~\bibnamefont
  {Garziano}}, \bibinfo {author} {\bibfnamefont {V.}~\bibnamefont
  {Macr{\`{i}}}}, \bibinfo {author} {\bibfnamefont {R.}~\bibnamefont {Stassi}},
  \bibinfo {author} {\bibfnamefont {O.}~\bibnamefont {{Di Stefano}}}, \bibinfo
  {author} {\bibfnamefont {F.}~\bibnamefont {Nori}},\ and\ \bibinfo {author}
  {\bibfnamefont {S.}~\bibnamefont {Savasta}},\ }\bibfield  {title} {\bibinfo
  {title} {{One Photon Can Simultaneously Excite Two or More Atoms}},\ }\href
  {https://doi.org/10.1103/PhysRevLett.117.043601} {\bibfield  {journal}
  {\bibinfo  {journal} {Physical Review Letters}\ }\textbf {\bibinfo {volume}
  {117}},\ \bibinfo {pages} {043601} (\bibinfo {year} {2016})}\BibitemShut
  {NoStop}%
\bibitem [{\citenamefont {Stassi}\ \emph {et~al.}(2017)\citenamefont {Stassi},
  \citenamefont {Macr{\`{i}}}, \citenamefont {Kockum}, \citenamefont {{Di
  Stefano}}, \citenamefont {Miranowicz}, \citenamefont {Savasta},\ and\
  \citenamefont {Nori}}]{Stassi2017}%
  \BibitemOpen
  \bibfield  {author} {\bibinfo {author} {\bibfnamefont {R.}~\bibnamefont
  {Stassi}}, \bibinfo {author} {\bibfnamefont {V.}~\bibnamefont {Macr{\`{i}}}},
  \bibinfo {author} {\bibfnamefont {A.~F.}\ \bibnamefont {Kockum}}, \bibinfo
  {author} {\bibfnamefont {O.}~\bibnamefont {{Di Stefano}}}, \bibinfo {author}
  {\bibfnamefont {A.}~\bibnamefont {Miranowicz}}, \bibinfo {author}
  {\bibfnamefont {S.}~\bibnamefont {Savasta}},\ and\ \bibinfo {author}
  {\bibfnamefont {F.}~\bibnamefont {Nori}},\ }\bibfield  {title} {\bibinfo
  {title} {{Quantum nonlinear optics without photons}},\ }\href
  {https://doi.org/10.1103/PhysRevA.96.023818} {\bibfield  {journal} {\bibinfo
  {journal} {Physical Review A}\ }\textbf {\bibinfo {volume} {96}},\ \bibinfo
  {pages} {023818} (\bibinfo {year} {2017})}\BibitemShut {NoStop}%
\bibitem [{\citenamefont {Kockum}\ \emph
  {et~al.}(2017{\natexlab{a}})\citenamefont {Kockum}, \citenamefont
  {Miranowicz}, \citenamefont {Macr{\`{i}}}, \citenamefont {Savasta},\ and\
  \citenamefont {Nori}}]{Kockum2017}%
  \BibitemOpen
  \bibfield  {author} {\bibinfo {author} {\bibfnamefont {A.~F.}\ \bibnamefont
  {Kockum}}, \bibinfo {author} {\bibfnamefont {A.}~\bibnamefont {Miranowicz}},
  \bibinfo {author} {\bibfnamefont {V.}~\bibnamefont {Macr{\`{i}}}}, \bibinfo
  {author} {\bibfnamefont {S.}~\bibnamefont {Savasta}},\ and\ \bibinfo {author}
  {\bibfnamefont {F.}~\bibnamefont {Nori}},\ }\bibfield  {title} {\bibinfo
  {title} {{Deterministic quantum nonlinear optics with single atoms and
  virtual photons}},\ }\href {https://doi.org/10.1103/PhysRevA.95.063849}
  {\bibfield  {journal} {\bibinfo  {journal} {Physical Review A}\ }\textbf
  {\bibinfo {volume} {95}},\ \bibinfo {pages} {063849} (\bibinfo {year}
  {2017}{\natexlab{a}})}\BibitemShut {NoStop}%
\bibitem [{\citenamefont {Kockum}\ \emph
  {et~al.}(2017{\natexlab{b}})\citenamefont {Kockum}, \citenamefont
  {Macr{\`{i}}}, \citenamefont {Garziano}, \citenamefont {Savasta},\ and\
  \citenamefont {Nori}}]{Kockum2017a}%
  \BibitemOpen
  \bibfield  {author} {\bibinfo {author} {\bibfnamefont {A.~F.}\ \bibnamefont
  {Kockum}}, \bibinfo {author} {\bibfnamefont {V.}~\bibnamefont {Macr{\`{i}}}},
  \bibinfo {author} {\bibfnamefont {L.}~\bibnamefont {Garziano}}, \bibinfo
  {author} {\bibfnamefont {S.}~\bibnamefont {Savasta}},\ and\ \bibinfo {author}
  {\bibfnamefont {F.}~\bibnamefont {Nori}},\ }\bibfield  {title} {\bibinfo
  {title} {{Frequency conversion in ultrastrong cavity QED}},\ }\href
  {https://doi.org/10.1038/s41598-017-04225-3} {\bibfield  {journal} {\bibinfo
  {journal} {Scientific Reports}\ }\textbf {\bibinfo {volume} {7}},\ \bibinfo
  {pages} {5313} (\bibinfo {year} {2017}{\natexlab{b}})}\BibitemShut {NoStop}%
\bibitem [{\citenamefont {Salmon}\ \emph {et~al.}(2022)\citenamefont {Salmon},
  \citenamefont {Gustin}, \citenamefont {Settineri}, \citenamefont {{Di
  Stefano}}, \citenamefont {Zueco}, \citenamefont {Savasta}, \citenamefont
  {Nori},\ and\ \citenamefont {Hughes}}]{Salmon2022}%
  \BibitemOpen
  \bibfield  {author} {\bibinfo {author} {\bibfnamefont {W.}~\bibnamefont
  {Salmon}}, \bibinfo {author} {\bibfnamefont {C.}~\bibnamefont {Gustin}},
  \bibinfo {author} {\bibfnamefont {A.}~\bibnamefont {Settineri}}, \bibinfo
  {author} {\bibfnamefont {O.}~\bibnamefont {{Di Stefano}}}, \bibinfo {author}
  {\bibfnamefont {D.}~\bibnamefont {Zueco}}, \bibinfo {author} {\bibfnamefont
  {S.}~\bibnamefont {Savasta}}, \bibinfo {author} {\bibfnamefont
  {F.}~\bibnamefont {Nori}},\ and\ \bibinfo {author} {\bibfnamefont
  {S.}~\bibnamefont {Hughes}},\ }\bibfield  {title} {\bibinfo {title}
  {{Gauge-independent emission spectra and quantum correlations in the
  ultrastrong coupling regime of open system cavity-QED}},\ }\href
  {https://doi.org/10.1515/nanoph-2021-0718} {\bibfield  {journal} {\bibinfo
  {journal} {Nanophotonics}\ }\textbf {\bibinfo {volume} {11}},\ \bibinfo
  {pages} {1573} (\bibinfo {year} {2022})}\BibitemShut {NoStop}%
\bibitem [{\citenamefont {Mercurio}\ \emph {et~al.}(2021)\citenamefont
  {Mercurio}, \citenamefont {Macr{\`{i}}}, \citenamefont {Gustin},
  \citenamefont {Hughes}, \citenamefont {Savasta},\ and\ \citenamefont
  {Nori}}]{Mercurio2022}%
  \BibitemOpen
  \bibfield  {author} {\bibinfo {author} {\bibfnamefont {A.}~\bibnamefont
  {Mercurio}}, \bibinfo {author} {\bibfnamefont {V.}~\bibnamefont
  {Macr{\`{i}}}}, \bibinfo {author} {\bibfnamefont {C.}~\bibnamefont {Gustin}},
  \bibinfo {author} {\bibfnamefont {S.}~\bibnamefont {Hughes}}, \bibinfo
  {author} {\bibfnamefont {S.}~\bibnamefont {Savasta}},\ and\ \bibinfo {author}
  {\bibfnamefont {F.}~\bibnamefont {Nori}},\ }\bibfield  {title} {\bibinfo
  {title} {{Regimes of Cavity-QED under Incoherent Excitation: From Weak to
  Deep Strong Coupling}},\ }\href
  {https://doi.org/10.1103/PHYSREVRESEARCH.4.023048/FIGURES/20/MEDIUM}
  {\bibfield  {journal} {\bibinfo  {journal} {Physical Review Research}\
  }\textbf {\bibinfo {volume} {4}},\ \bibinfo {pages} {023048} (\bibinfo {year}
  {2021})}\BibitemShut {NoStop}%
\bibitem [{\citenamefont {del Valle}\ \emph {et~al.}(2012)\citenamefont {del
  Valle}, \citenamefont {Gonzalez-Tudela}, \citenamefont {Laussy},
  \citenamefont {Tejedor},\ and\ \citenamefont {Hartmann}}]{DelValle2012}%
  \BibitemOpen
  \bibfield  {author} {\bibinfo {author} {\bibfnamefont {E.}~\bibnamefont {del
  Valle}}, \bibinfo {author} {\bibfnamefont {A.}~\bibnamefont
  {Gonzalez-Tudela}}, \bibinfo {author} {\bibfnamefont {F.~P.}\ \bibnamefont
  {Laussy}}, \bibinfo {author} {\bibfnamefont {C.}~\bibnamefont {Tejedor}},\
  and\ \bibinfo {author} {\bibfnamefont {M.~J.}\ \bibnamefont {Hartmann}},\
  }\bibfield  {title} {\bibinfo {title} {{Theory of Frequency-Filtered and
  Time-Resolved $N$-Photon Correlations}},\ }\href
  {https://doi.org/10.1103/PhysRevLett.109.183601} {\bibfield  {journal}
  {\bibinfo  {journal} {Physical Review Letters}\ }\textbf {\bibinfo {volume}
  {109}},\ \bibinfo {pages} {183601} (\bibinfo {year} {2012})}\BibitemShut
  {NoStop}%
\bibitem [{\citenamefont {Salmon}(2021)}]{Salmon2021}%
  \BibitemOpen
  \bibfield  {author} {\bibinfo {author} {\bibfnamefont {W.}~\bibnamefont
  {Salmon}},\ }\emph {\bibinfo {title} {{Master Equations for Computing
  Gauge-Invariant Observables in the Ultrastrong Coupling Regime of
  Cavity-QED}}},\ \href@noop {} {\bibinfo {type} {Doctoral dissertation}},\
  \bibinfo  {school} {Queen's University (Canada)} (\bibinfo {year}
  {2021})\BibitemShut {NoStop}%
\bibitem [{\citenamefont {Chestnov}\ \emph {et~al.}(2017)\citenamefont
  {Chestnov}, \citenamefont {Shahnazaryan}, \citenamefont {Alodjants},\ and\
  \citenamefont {Shelykh}}]{Chestnov2017}%
  \BibitemOpen
  \bibfield  {author} {\bibinfo {author} {\bibfnamefont {I.~Y.}\ \bibnamefont
  {Chestnov}}, \bibinfo {author} {\bibfnamefont {V.~A.}\ \bibnamefont
  {Shahnazaryan}}, \bibinfo {author} {\bibfnamefont {A.~P.}\ \bibnamefont
  {Alodjants}},\ and\ \bibinfo {author} {\bibfnamefont {I.~A.}\ \bibnamefont
  {Shelykh}},\ }\bibfield  {title} {\bibinfo {title} {{Terahertz Lasing in
  Ensemble of Asymmetric Quantum Dots}},\ }\href
  {https://doi.org/10.1021/acsphotonics.7b00575} {\bibfield  {journal}
  {\bibinfo  {journal} {ACS Photonics}\ }\textbf {\bibinfo {volume} {4}},\
  \bibinfo {pages} {2726} (\bibinfo {year} {2017})}\BibitemShut {NoStop}%
\bibitem [{\citenamefont {Settineri}\ \emph {et~al.}(2021)\citenamefont
  {Settineri}, \citenamefont {{Di Stefano}}, \citenamefont {Zueco},
  \citenamefont {Hughes}, \citenamefont {Savasta},\ and\ \citenamefont
  {Nori}}]{Settineri2021}%
  \BibitemOpen
  \bibfield  {author} {\bibinfo {author} {\bibfnamefont {A.}~\bibnamefont
  {Settineri}}, \bibinfo {author} {\bibfnamefont {O.}~\bibnamefont {{Di
  Stefano}}}, \bibinfo {author} {\bibfnamefont {D.}~\bibnamefont {Zueco}},
  \bibinfo {author} {\bibfnamefont {S.}~\bibnamefont {Hughes}}, \bibinfo
  {author} {\bibfnamefont {S.}~\bibnamefont {Savasta}},\ and\ \bibinfo {author}
  {\bibfnamefont {F.}~\bibnamefont {Nori}},\ }\bibfield  {title} {\bibinfo
  {title} {{Gauge freedom, quantum measurements, and time-dependent
  interactions in cavity QED}},\ }\href
  {https://doi.org/10.1103/PhysRevResearch.3.023079} {\bibfield  {journal}
  {\bibinfo  {journal} {Physical Review Research}\ }\textbf {\bibinfo {volume}
  {3}},\ \bibinfo {pages} {023079} (\bibinfo {year} {2021})}\BibitemShut
  {NoStop}%
\bibitem [{\citenamefont {{De Bernardis}}\ \emph {et~al.}(2018)\citenamefont
  {{De Bernardis}}, \citenamefont {Pilar}, \citenamefont {Jaako}, \citenamefont
  {{De Liberato}},\ and\ \citenamefont {Rabl}}]{DeBernardis2018}%
  \BibitemOpen
  \bibfield  {author} {\bibinfo {author} {\bibfnamefont {D.}~\bibnamefont {{De
  Bernardis}}}, \bibinfo {author} {\bibfnamefont {P.}~\bibnamefont {Pilar}},
  \bibinfo {author} {\bibfnamefont {T.}~\bibnamefont {Jaako}}, \bibinfo
  {author} {\bibfnamefont {S.}~\bibnamefont {{De Liberato}}},\ and\ \bibinfo
  {author} {\bibfnamefont {P.}~\bibnamefont {Rabl}},\ }\bibfield  {title}
  {\bibinfo {title} {{Breakdown of gauge invariance in ultrastrong-coupling
  cavity QED}},\ }\href {https://doi.org/10.1103/PhysRevA.98.053819} {\bibfield
   {journal} {\bibinfo  {journal} {Physical Review A}\ }\textbf {\bibinfo
  {volume} {98}},\ \bibinfo {pages} {053819} (\bibinfo {year}
  {2018})}\BibitemShut {NoStop}%
\bibitem [{\citenamefont {{Di Stefano}}\ \emph {et~al.}(2019)\citenamefont {{Di
  Stefano}}, \citenamefont {Settineri}, \citenamefont {Macr{\`{i}}},
  \citenamefont {Garziano}, \citenamefont {Stassi}, \citenamefont {Savasta},\
  and\ \citenamefont {Nori}}]{DiStefano2019}%
  \BibitemOpen
  \bibfield  {author} {\bibinfo {author} {\bibfnamefont {O.}~\bibnamefont {{Di
  Stefano}}}, \bibinfo {author} {\bibfnamefont {A.}~\bibnamefont {Settineri}},
  \bibinfo {author} {\bibfnamefont {V.}~\bibnamefont {Macr{\`{i}}}}, \bibinfo
  {author} {\bibfnamefont {L.}~\bibnamefont {Garziano}}, \bibinfo {author}
  {\bibfnamefont {R.}~\bibnamefont {Stassi}}, \bibinfo {author} {\bibfnamefont
  {S.}~\bibnamefont {Savasta}},\ and\ \bibinfo {author} {\bibfnamefont
  {F.}~\bibnamefont {Nori}},\ }\bibfield  {title} {\bibinfo {title}
  {{Resolution of gauge ambiguities in ultrastrong-coupling cavity quantum
  electrodynamics}},\ }\href {https://doi.org/10.1038/s41567-019-0534-4}
  {\bibfield  {journal} {\bibinfo  {journal} {Nature Physics}\ }\textbf
  {\bibinfo {volume} {15}},\ \bibinfo {pages} {803} (\bibinfo {year}
  {2019})}\BibitemShut {NoStop}%
\bibitem [{Note1()}]{Note1}%
  \BibitemOpen
  \bibinfo {note} {See Supplemental Material for further details on the
  derivation of the Quantum Rabi Model Hamiltonian in the dipole gauge and on
  the calculation of time-averaged density matrix form time-dependent
  Liouvillians using the Floquet method. Includes references~\cite
  {Settineri2021,Salmon2022,DeBernardis2018,DiStefano2019,Majumdar2011g,Papageorge2012a,Maragkou2013}.}\BibitemShut
  {Stop}%
\bibitem [{\citenamefont {Ridolfo}\ \emph {et~al.}(2012)\citenamefont
  {Ridolfo}, \citenamefont {Leib}, \citenamefont {Savasta},\ and\ \citenamefont
  {Hartmann}}]{Ridolfo2012}%
  \BibitemOpen
  \bibfield  {author} {\bibinfo {author} {\bibfnamefont {A.}~\bibnamefont
  {Ridolfo}}, \bibinfo {author} {\bibfnamefont {M.}~\bibnamefont {Leib}},
  \bibinfo {author} {\bibfnamefont {S.}~\bibnamefont {Savasta}},\ and\ \bibinfo
  {author} {\bibfnamefont {M.~J.}\ \bibnamefont {Hartmann}},\ }\bibfield
  {title} {\bibinfo {title} {{Photon Blockade in the Ultrastrong Coupling
  Regime}},\ }\href {https://doi.org/10.1103/PhysRevLett.109.193602} {\bibfield
   {journal} {\bibinfo  {journal} {Physical Review Letters}\ }\textbf {\bibinfo
  {volume} {109}},\ \bibinfo {pages} {193602} (\bibinfo {year}
  {2012})}\BibitemShut {NoStop}%
\bibitem [{\citenamefont {Beaudoin}\ \emph {et~al.}(2011)\citenamefont
  {Beaudoin}, \citenamefont {Gambetta},\ and\ \citenamefont
  {Blais}}]{Beaudoin2011}%
  \BibitemOpen
  \bibfield  {author} {\bibinfo {author} {\bibfnamefont {F.}~\bibnamefont
  {Beaudoin}}, \bibinfo {author} {\bibfnamefont {J.~M.}\ \bibnamefont
  {Gambetta}},\ and\ \bibinfo {author} {\bibfnamefont {A.}~\bibnamefont
  {Blais}},\ }\bibfield  {title} {\bibinfo {title} {{Dissipation and
  ultrastrong coupling in circuit QED}},\ }\href
  {https://doi.org/10.1103/PhysRevA.84.043832} {\bibfield  {journal} {\bibinfo
  {journal} {Physical Review A}\ }\textbf {\bibinfo {volume} {84}},\ \bibinfo
  {pages} {043832} (\bibinfo {year} {2011})}\BibitemShut {NoStop}%
\bibitem [{\citenamefont {{Di Stefano}}\ \emph {et~al.}(2018)\citenamefont {{Di
  Stefano}}, \citenamefont {Kockum}, \citenamefont {Ridolfo}, \citenamefont
  {Savasta},\ and\ \citenamefont {Nori}}]{DiStefano2018}%
  \BibitemOpen
  \bibfield  {author} {\bibinfo {author} {\bibfnamefont {O.}~\bibnamefont {{Di
  Stefano}}}, \bibinfo {author} {\bibfnamefont {A.~F.}\ \bibnamefont {Kockum}},
  \bibinfo {author} {\bibfnamefont {A.}~\bibnamefont {Ridolfo}}, \bibinfo
  {author} {\bibfnamefont {S.}~\bibnamefont {Savasta}},\ and\ \bibinfo {author}
  {\bibfnamefont {F.}~\bibnamefont {Nori}},\ }\bibfield  {title} {\bibinfo
  {title} {{Photodetection probability in quantum systems with arbitrarily
  strong light-matter interaction}},\ }\href
  {https://doi.org/10.1038/s41598-018-36056-1} {\bibfield  {journal} {\bibinfo
  {journal} {Scientific Reports}\ }\textbf {\bibinfo {volume} {8}},\ \bibinfo
  {pages} {17825} (\bibinfo {year} {2018})}\BibitemShut {NoStop}%
\bibitem [{\citenamefont {{Le Boit{\'{e}}}}(2020)}]{LeBoite2020}%
  \BibitemOpen
  \bibfield  {author} {\bibinfo {author} {\bibfnamefont {A.}~\bibnamefont {{Le
  Boit{\'{e}}}}},\ }\bibfield  {title} {\bibinfo {title} {{Theoretical Methods
  for Ultrastrong Light–Matter Interactions}},\ }\href
  {https://doi.org/10.1002/qute.201900140} {\bibfield  {journal} {\bibinfo
  {journal} {Advanced Quantum Technologies}\ }\textbf {\bibinfo {volume} {3}},\
  \bibinfo {pages} {1900140} (\bibinfo {year} {2020})}\BibitemShut {NoStop}%
\bibitem [{\citenamefont {Settineri}\ \emph {et~al.}(2018)\citenamefont
  {Settineri}, \citenamefont {Macr{\'{i}}}, \citenamefont {Ridolfo},
  \citenamefont {{Di Stefano}}, \citenamefont {Kockum}, \citenamefont {Nori},\
  and\ \citenamefont {Savasta}}]{Settineri2018}%
  \BibitemOpen
  \bibfield  {author} {\bibinfo {author} {\bibfnamefont {A.}~\bibnamefont
  {Settineri}}, \bibinfo {author} {\bibfnamefont {V.}~\bibnamefont
  {Macr{\'{i}}}}, \bibinfo {author} {\bibfnamefont {A.}~\bibnamefont
  {Ridolfo}}, \bibinfo {author} {\bibfnamefont {O.}~\bibnamefont {{Di
  Stefano}}}, \bibinfo {author} {\bibfnamefont {A.~F.}\ \bibnamefont {Kockum}},
  \bibinfo {author} {\bibfnamefont {F.}~\bibnamefont {Nori}},\ and\ \bibinfo
  {author} {\bibfnamefont {S.}~\bibnamefont {Savasta}},\ }\bibfield  {title}
  {\bibinfo {title} {{Dissipation and thermal noise in hybrid quantum systems
  in the ultrastrong-coupling regime}},\ }\href
  {https://doi.org/10.1103/PhysRevA.98.053834} {\bibfield  {journal} {\bibinfo
  {journal} {Physical Review A}\ }\textbf {\bibinfo {volume} {98}},\ \bibinfo
  {pages} {053834} (\bibinfo {year} {2018})}\BibitemShut {NoStop}%
\bibitem [{\citenamefont {Faraon}\ \emph {et~al.}(2008)\citenamefont {Faraon},
  \citenamefont {Fushman}, \citenamefont {Englund}, \citenamefont {Stoltz},
  \citenamefont {Petroff},\ and\ \citenamefont
  {Vu{\v{c}}kovi{\'{c}}}}]{Faraon2008}%
  \BibitemOpen
  \bibfield  {author} {\bibinfo {author} {\bibfnamefont {A.}~\bibnamefont
  {Faraon}}, \bibinfo {author} {\bibfnamefont {I.}~\bibnamefont {Fushman}},
  \bibinfo {author} {\bibfnamefont {D.}~\bibnamefont {Englund}}, \bibinfo
  {author} {\bibfnamefont {N.}~\bibnamefont {Stoltz}}, \bibinfo {author}
  {\bibfnamefont {P.}~\bibnamefont {Petroff}},\ and\ \bibinfo {author}
  {\bibfnamefont {J.}~\bibnamefont {Vu{\v{c}}kovi{\'{c}}}},\ }\bibfield
  {title} {\bibinfo {title} {{Coherent generation of non-classical light on a
  chip via photon-induced tunnelling and blockade}},\ }\href
  {https://doi.org/10.1038/nphys1078} {\bibfield  {journal} {\bibinfo
  {journal} {Nature Physics}\ }\textbf {\bibinfo {volume} {4}},\ \bibinfo
  {pages} {859} (\bibinfo {year} {2008})}\BibitemShut {NoStop}%
\bibitem [{\citenamefont {Chang}\ \emph {et~al.}(2014)\citenamefont {Chang},
  \citenamefont {Vuleti{\'{c}}},\ and\ \citenamefont {Lukin}}]{Chang2014a}%
  \BibitemOpen
  \bibfield  {author} {\bibinfo {author} {\bibfnamefont {D.~E.}\ \bibnamefont
  {Chang}}, \bibinfo {author} {\bibfnamefont {V.}~\bibnamefont
  {Vuleti{\'{c}}}},\ and\ \bibinfo {author} {\bibfnamefont {M.~D.}\
  \bibnamefont {Lukin}},\ }\bibfield  {title} {\bibinfo {title} {{Quantum
  nonlinear optics — photon by photon}},\ }\href
  {https://doi.org/10.1038/nphoton.2014.192} {\bibfield  {journal} {\bibinfo
  {journal} {Nature Photonics}\ }\textbf {\bibinfo {volume} {8}},\ \bibinfo
  {pages} {685} (\bibinfo {year} {2014})}\BibitemShut {NoStop}%
\bibitem [{\citenamefont {M{\"{u}}ller}\ \emph {et~al.}(2015)\citenamefont
  {M{\"{u}}ller}, \citenamefont {Rundquist}, \citenamefont {Fischer},
  \citenamefont {Sarmiento}, \citenamefont {Lagoudakis}, \citenamefont
  {Kelaita}, \citenamefont {{S{\'{a}}nchez Mu{\~{n}}oz}}, \citenamefont {del
  Valle}, \citenamefont {Laussy},\ and\ \citenamefont
  {Vu{\v{c}}kovi{\'{c}}}}]{Muller2015a}%
  \BibitemOpen
  \bibfield  {author} {\bibinfo {author} {\bibfnamefont {K.}~\bibnamefont
  {M{\"{u}}ller}}, \bibinfo {author} {\bibfnamefont {A.}~\bibnamefont
  {Rundquist}}, \bibinfo {author} {\bibfnamefont {K.~A.}\ \bibnamefont
  {Fischer}}, \bibinfo {author} {\bibfnamefont {T.}~\bibnamefont {Sarmiento}},
  \bibinfo {author} {\bibfnamefont {K.~G.}\ \bibnamefont {Lagoudakis}},
  \bibinfo {author} {\bibfnamefont {Y.~A.}\ \bibnamefont {Kelaita}}, \bibinfo
  {author} {\bibfnamefont {C.}~\bibnamefont {{S{\'{a}}nchez Mu{\~{n}}oz}}},
  \bibinfo {author} {\bibfnamefont {E.}~\bibnamefont {del Valle}}, \bibinfo
  {author} {\bibfnamefont {F.~P.}\ \bibnamefont {Laussy}},\ and\ \bibinfo
  {author} {\bibfnamefont {J.}~\bibnamefont {Vu{\v{c}}kovi{\'{c}}}},\
  }\bibfield  {title} {\bibinfo {title} {{Coherent Generation of Nonclassical
  Light on Chip via Detuned Photon Blockade}},\ }\href
  {https://doi.org/10.1103/PhysRevLett.114.233601} {\bibfield  {journal}
  {\bibinfo  {journal} {Physical Review Letters}\ }\textbf {\bibinfo {volume}
  {114}},\ \bibinfo {pages} {233601} (\bibinfo {year} {2015})}\BibitemShut
  {NoStop}%
\bibitem [{\citenamefont {Hamsen}\ \emph {et~al.}(2017)\citenamefont {Hamsen},
  \citenamefont {Tolazzi}, \citenamefont {Wilk},\ and\ \citenamefont
  {Rempe}}]{Hamsen2017}%
  \BibitemOpen
  \bibfield  {author} {\bibinfo {author} {\bibfnamefont {C.}~\bibnamefont
  {Hamsen}}, \bibinfo {author} {\bibfnamefont {K.~N.}\ \bibnamefont {Tolazzi}},
  \bibinfo {author} {\bibfnamefont {T.}~\bibnamefont {Wilk}},\ and\ \bibinfo
  {author} {\bibfnamefont {G.}~\bibnamefont {Rempe}},\ }\bibfield  {title}
  {\bibinfo {title} {{Two-Photon Blockade in an Atom-Driven Cavity QED
  System}},\ }\href {https://doi.org/10.1103/PhysRevLett.118.133604} {\bibfield
   {journal} {\bibinfo  {journal} {Physical Review Letters}\ }\textbf {\bibinfo
  {volume} {118}},\ \bibinfo {pages} {133604} (\bibinfo {year}
  {2017})}\BibitemShut {NoStop}%
\bibitem [{\citenamefont {Majumdar}\ \emph {et~al.}(2011)\citenamefont
  {Majumdar}, \citenamefont {Papageorge}, \citenamefont {Kim}, \citenamefont
  {Bajcsy}, \citenamefont {Kim}, \citenamefont {Petroff},\ and\ \citenamefont
  {Vu{\v{c}}kovi{\'{c}}}}]{Majumdar2011g}%
  \BibitemOpen
  \bibfield  {author} {\bibinfo {author} {\bibfnamefont {A.}~\bibnamefont
  {Majumdar}}, \bibinfo {author} {\bibfnamefont {A.}~\bibnamefont
  {Papageorge}}, \bibinfo {author} {\bibfnamefont {E.~D.}\ \bibnamefont {Kim}},
  \bibinfo {author} {\bibfnamefont {M.}~\bibnamefont {Bajcsy}}, \bibinfo
  {author} {\bibfnamefont {H.}~\bibnamefont {Kim}}, \bibinfo {author}
  {\bibfnamefont {P.}~\bibnamefont {Petroff}},\ and\ \bibinfo {author}
  {\bibfnamefont {J.}~\bibnamefont {Vu{\v{c}}kovi{\'{c}}}},\ }\bibfield
  {title} {\bibinfo {title} {{Probing of single quantum dot dressed states via
  an off-resonant cavity}},\ }\href
  {https://doi.org/10.1103/PhysRevB.84.085310} {\bibfield  {journal} {\bibinfo
  {journal} {Physical Review B}\ }\textbf {\bibinfo {volume} {84}},\ \bibinfo
  {pages} {085310} (\bibinfo {year} {2011})}\BibitemShut {NoStop}%
\bibitem [{\citenamefont {Papageorge}\ \emph {et~al.}(2012)\citenamefont
  {Papageorge}, \citenamefont {Majumdar}, \citenamefont {Kim},\ and\
  \citenamefont {Vu{\v{c}}kovi{\'{c}}}}]{Papageorge2012a}%
  \BibitemOpen
  \bibfield  {author} {\bibinfo {author} {\bibfnamefont {A.}~\bibnamefont
  {Papageorge}}, \bibinfo {author} {\bibfnamefont {A.}~\bibnamefont
  {Majumdar}}, \bibinfo {author} {\bibfnamefont {E.~D.}\ \bibnamefont {Kim}},\
  and\ \bibinfo {author} {\bibfnamefont {J.}~\bibnamefont
  {Vu{\v{c}}kovi{\'{c}}}},\ }\bibfield  {title} {\bibinfo {title} {{Bichromatic
  driving of a solid-state cavity quantum electrodynamics system}},\ }\href
  {https://doi.org/10.1088/1367-2630/14/1/013028} {\bibfield  {journal}
  {\bibinfo  {journal} {New Journal of Physics}\ }\textbf {\bibinfo {volume}
  {14}},\ \bibinfo {pages} {013028} (\bibinfo {year} {2012})}\BibitemShut
  {NoStop}%
\bibitem [{\citenamefont {Maragkou}\ \emph {et~al.}(2013)\citenamefont
  {Maragkou}, \citenamefont {S{\'{a}}nchez-Mu{\~{n}}oz}, \citenamefont
  {Lazi{\'{c}}}, \citenamefont {Chernysheva}, \citenamefont {van~der Meulen},
  \citenamefont {Gonz{\'{a}}lez-Tudela}, \citenamefont {Tejedor}, \citenamefont
  {Mart{\'{i}}nez}, \citenamefont {Prieto}, \citenamefont {Postigo},\ and\
  \citenamefont {Calleja}}]{Maragkou2013}%
  \BibitemOpen
  \bibfield  {author} {\bibinfo {author} {\bibfnamefont {M.}~\bibnamefont
  {Maragkou}}, \bibinfo {author} {\bibfnamefont {C.}~\bibnamefont
  {S{\'{a}}nchez-Mu{\~{n}}oz}}, \bibinfo {author} {\bibfnamefont
  {S.}~\bibnamefont {Lazi{\'{c}}}}, \bibinfo {author} {\bibfnamefont
  {E.}~\bibnamefont {Chernysheva}}, \bibinfo {author} {\bibfnamefont {H.~P.}\
  \bibnamefont {van~der Meulen}}, \bibinfo {author} {\bibfnamefont
  {A.}~\bibnamefont {Gonz{\'{a}}lez-Tudela}}, \bibinfo {author} {\bibfnamefont
  {C.}~\bibnamefont {Tejedor}}, \bibinfo {author} {\bibfnamefont {L.~J.}\
  \bibnamefont {Mart{\'{i}}nez}}, \bibinfo {author} {\bibfnamefont
  {I.}~\bibnamefont {Prieto}}, \bibinfo {author} {\bibfnamefont {P.~A.}\
  \bibnamefont {Postigo}},\ and\ \bibinfo {author} {\bibfnamefont {J.~M.}\
  \bibnamefont {Calleja}},\ }\bibfield  {title} {\bibinfo {title} {{Bichromatic
  dressing of a quantum dot detected by a remote second quantum dot}},\ }\href
  {https://doi.org/10.1103/PhysRevB.88.075309} {\bibfield  {journal} {\bibinfo
  {journal} {Physical Review B}\ }\textbf {\bibinfo {volume} {88}},\ \bibinfo
  {pages} {075309} (\bibinfo {year} {2013})}\BibitemShut {NoStop}%
\end{thebibliography}%

\clearpage
\onecolumngrid

\begin{center}
{\bf \large Supplementary Material}
\end{center}

\renewcommand{\theequation}{S\arabic{equation}}

\renewcommand{\thefigure}{S\arabic{figure}} 
\setcounter{figure}{0} 
\setcounter{equation}{0}

\section{Hamiltonian in the dipole gauge}

In the dipole gauge, the field conjugate momentum corresponds to the displacement operator $\hat{\bf \Pi}({\bf r}) = - \hat{\bf D}({\bf r})  =- i \epsilon_0 {\bf E}_0({\bf r}) (\aop - \adop)$, where ${\bf E}_0({\bf r}) = \sqrt\frac{\omega_c}{2\epsilon_0}{\bf f}_0 ({\bf r})$, with ${\bf f}_0 ({\bf r})$ the mode function of the cavity mode under consideration. In this gauge, the electric field operator is not described only by photon operators; taking into account that $\epsilon_0\hat{\bf E} = \hat{\bf D} - \hat{\bf P}$, respectively, one can prove~\cite{Settineri2021} that $\bfop E$ adopts the form $\bfop E \approx \mathbf{E}_0(\mathbf r)[i(\aop-\adop) - 2\eta \spol]$, where $\eta \equiv {\bm \mu} \cdot {\bf E}_0({\bf r}_0)/\omega_c $ is the dimensionless coupling parameter between cavity and TLS, and we have neglected the small contribution of the polarizability of the ancilla sensor. In the dipole gauge, the interaction Hamiltonian thus takes the form~\cite{Settineri2021,Salmon2022}:
\begin{equation}
\ham_I  = i\eta \omega_c (\adop - \aop)\spol + \omega_c\eta_s[i(\adop-\aop)+2\eta\spol]\sx^s.
\label{eq:H_int}
\end{equation}
The choice of this gauge ensures a proper gauge invariance under the TLS approximation \cite{DeBernardis2018,DiStefano2019,Salmon2022}. Here, $\eta_s \equiv {\bm \mu}_s \cdot {\bf E}_0({\bf r}_0)/\omega_c $  is the coupling parameter between sensor and cavity.  In this work we will focus on phenomenology emerging on the ultrastrong coupling regime, where $\eta\sim 0.1$, and fix $\eta_s \lll 1$ so that the sensor qubit only probes the dynamics of the TLS-cavity system, without altering it.
From now on, we consider that the problem is confined to a single polarization, i.e., all vector quantities are aligned along the same unit vector ${\bf u}_x,$ and substitute vector field operators $\bfop E$ by scalar operators, so that $\bfop{E}=\hat E {\bf u}_x$. Finally, the drive Hamiltonian reads $\ham_\text{drive}(t) =\Omega[i(\aop - \adop)-2\eta\sx]\cos(\omega_\mathrm L t)$, 
so that $\ham_\text{drive} \propto \hat E \cos(\omega_\mathrm L t)$.

\section{Averaged stationary state of a time-dependent problem: Floquet theory}

Due to the presence of counter-rotating terms in the Hamiltonian
\begin{equation}
\ham  = i\eta \omega_c (\adop - \aop)\spol + \omega_c\eta_s[i(\adop-\aop)+2\eta\spol]\sx^s +\Omega[i(\aop - \adop)-2\eta\sx]\cos(\omega_\mathrm L t),
\label{eq:H_int}
\end{equation}
one cannot perform the standard unitary transformation to the rotating frame of the drive that leads to a time-independent Hamiltonian, and thus we are left with the explicitly time-dependent terms. The Liouvillian superoperator (defined as the generator of the master equation $\dot\rho = \mathcal L \rho$) can then be cast as $\mathcal L = \mathcal L_0 + \mathcal L_+ e^{i\omega_\mathrm L t}+ \mathcal L_- e^{-i\omega_\mathrm L t}$. As a consequence, there is not a fully stationary state in the long-time limit, 
 but a time-dependent state whose form we can postulate as ${\rho(t\rightarrow \infty) = \sum_{n=-\infty}^\infty \rho_n e^{i n\omega_\mathrm L t}}$, which allows us to solve the problem using Floquet theory~\cite{Majumdar2011g,Papageorge2012a,Maragkou2013}. The time-averaged steady-state matrix is $\rho^\mathrm{ss} = \rho_0$, and can be found as the nullspace of $\mathcal L_0 + \mathcal L _- \mathcal S_1 + \mathcal L _+ \mathcal S_{-1}$, where the $\mathcal S_n$ operators are obtained recursively as
\begin{equation}
\mathcal S_{\pm n} = -[\mathcal L_0  - (z+ i n \omega_\mathrm{L}) + \mathcal L _{\mp} \mathcal S_{\pm(n+1)}]^{-1}\mathcal L_\pm,
\end{equation}
which is solved assuming that $\mathcal S_{\pm n_\mathrm{max}} = 0$ for a sufficiently large $n_\mathrm{max}$. Once a time-averaged steady state is obtained, we consider that the stationary rate of emission from the sensor is proportional to the spectrum of emission at the sensor's frequency, i.e. $S(\omega_s) \propto \text{Tr}[\rho_{\mathrm ss}\Sigma^-_s \hat\Sigma_s^+]$, with the sensor's decay rate $\Gamma$ corresponding to the filter linewidth. One is thus able to compute the spectrum by scanning the sensor's frequency and computing the corresponding  time-averaged steady state.

\end{document}